\RequirePackage{fix-cm}

\documentclass[smallextended]{svjour3}       
\smartqed  

\usepackage{textgreek}
\usepackage{array}
\usepackage{mathptmx}
\usepackage{graphicx}
\usepackage{listings}
\usepackage{verbatimbox}
\usepackage{subfig}
\usepackage{paralist}
\usepackage{xspace}
\usepackage{booktabs}
\usepackage{color}
\usepackage{setspace}
\usepackage{url}
\usepackage[numbers, sort&compress]{natbib}
\usepackage[colorlinks,bookmarksopen,bookmarksnumbered,citecolor=red,urlcolor=red]{hyperref}
\usepackage{flushend}
\usepackage{subfig}
\usepackage{color}
\usepackage{float}
\usepackage{amsmath}
\usepackage{amssymb}
\usepackage{fancybox}
\usepackage{rotating}
\usepackage{tablefootnote}
\usepackage[para,online,flushleft]{threeparttable}
\usepackage[referable]{threeparttablex}

\newcommand{\hypobox}[1]{\begin{center}%
	\noindent\thicklines\setlength{\fboxsep}{7pt}%
	\cornersize{0}\Ovalbox{\begin{minipage}{0.9\textwidth}%
	\vspace{-0.1cm}
	\textit{#1}
	\vspace{-0.1cm}
	\end{minipage}} \end{center}}

\newcommand{\code}[1]{{\texttt{#1}}}

\definecolor{mygray}{gray}{0.4}
\newfloat{Code}{H}{myc}

\begin{document}

\title{Building the Perfect Game -- An Empirical Study of Game Modifications}
\author{
%
%
\alignauthor
Daniel Lee\\
       \affaddr{Software Analysis and Intelligence Lab (SAIL)}\\
       \affaddr{Queen's University}\\
       \affaddr{Kingston, ON, Canada}\\
       \email{dlee@cs.queensu.ca}
\alignauthor
Dayi Lin\\
       \affaddr{Software Analysis and Intelligence Lab (SAIL)}\\
       \affaddr{Queen's University}\\
       \affaddr{Kingston, ON, Canada}\\
       \email{dayi.lin@cs.queensu.ca}    
\alignauthor
Cor-Paul Bezemer\\
       \affaddr{Analytics of Software, Games and Repository Data (ASGAARD) Lab}\\
       \affaddr{University of Alberta}\\
       \affaddr{Edmonton, AB, Canada}\\
       \email{bezemer@ualberta.ca}
\alignauthor
Ahmed E. Hassan\\
       \affaddr{Software Analysis and Intelligence Lab (SAIL)}\\
       \affaddr{Queen's University}\\
       \affaddr{Kingston, ON, Canada}\\
       \email{ahmed@cs.queensu.ca}
}

\author{Daniel Lee \and Dayi Lin \and Cor-Paul Bezemer \and Ahmed E. Hassan }


\institute{Daniel Lee \and Dayi Lin \and Ahmed E. Hassan  \at
              Software Analysis and Intelligence Lab (SAIL)\\
              Queen's University\\
              Kingston, ON, Canada \\
              \email{\{dlee, dayi.lin, ahmed\}@cs.queensu.ca}  \\ \\
Cor-Paul Bezemer \at
              Analytics of Software, Games and Repository Data (ASGAARD) Lab\\
              University of Alberta\\
              Edmonton, AB, Canada \\
              \email{bezemer@ualberta.ca}  \\
}

\date{Received: date / Accepted: date}

\maketitle

\begin{abstract} Prior work has shown that gamer loyalty is important for the sales of a developer's future games. Therefore, it is important for game developers to increase the longevity of their games. However, game developers cannot always meet the growing and changing needs of the gaming community, due to the often already overloaded schedules of developers. So-called modders can potentially assist game developers with addressing gamers' needs. Modders are enthusiasts who provide modifications or completely new content for a game.
By supporting modders, game developers can meet the rapidly growing and varying needs of their gamer base. Modders have the potential to play a role in extending the life expectancy of a game, thereby saving game developers time and money, and leading to a better overall gaming experience for their gamer base.

In this paper, we empirically study the metadata of 9,521 mods that were extracted from the Nexus Mods distribution platform. The Nexus Mods distribution platform is one of the largest mod distribution platforms for PC games at the time of our study. The goal of our paper is to provide useful insights about mods on the Nexus Mods distribution platform from a quantitative perspective, and to provide researchers a solid foundation to further explore game mods. To better understand the potential of mods to extend the longevity of a game we study their characteristics, and we study their release schedules and post-release support (in terms of bug reports) as a proxy for the willingness of the modding community to contribute to a game. We find that providing official support for mods can be beneficial for the perceived quality of the mods of a game: games for which a modding tool is provided by the original game developer have a higher median endorsement ratio than mods for games that do not have such a tool. 
In addition, mod users are willing to submit bug reports for a mod. However, they often fail to do this in a systematic manner using the bug reporting tool of the Nexus Mods platform, resulting in low-quality bug reports which are difficult to resolve.

Our findings give the first insights into the characteristics, release schedule and post-release support of game mods. Our findings show that some games have a very active modding community, which contributes to those games through mods. Based on our findings, we recommend that game developers who desire an active modding community for their own games provide the modding community with an officially-supported modding tool. In addition, we recommend that mod distribution platforms, such as Nexus Mods, improve their bug reporting system to receive higher quality bug reports.

\end{abstract}


\section{Introduction}
\label{sec:intro}

The gaming industry keeps on growing at a rapid pace. The Global Games Market projected that 2.3 billion gamers would spend \$137.9 billion USD on digital games in 2018, representing a 13\% increase from 2017~\cite{globalgames2018}. 
Because of the scale of the gaming industry, and the fact that gamers are
extremely hard to please~\cite{chambers2005measurement}, it has become challenging, and often costly, to develop a successful game.
One strategy that has proven to be profitable for game developers is to increase gamer retention, e.g., by striving to keep current players playing a game.
Prior work~\cite{gamerloyalty} shows that long time-players of a game are more likely to buy the future games of the same developer. Hence, gamer loyalty is an important aspect for game developers.

However, keeping up with the changing needs of gamers is hard, which makes maintaining gamer loyalty difficult as well. The increasing gamer expectations and rising development cost are increasing the pressure on game development teams~\cite{sechallenges}. Developers need to balance their available resources between working on a new(er) game, maintenance updates of an older game, and updates (such as new game levels) that are desired by the players of that game. Hence, the tradeoff between increasing the longevity of an older game and working on a newer game is a challenging tradeoff for game developers.

For some games the gaming community itself extends a helping hand with increasing the longevity of the game.
Game modding is the practice of modifying an existing game through \emph{game mods}~\cite{modopensource}. Game modders are external developers or game enthusiasts who make modifications to an existing game, because they enjoy or want to improve the original game. 
Game mods allow an original game to remain replayable for longer than it was originally intended to, as the mods add new or updated content to the game.
For example, the \textit{Skyrim} game\footnote{\url{https://elderscrolls.bethesda.net/en/skyrim}} continues to receive new content six years after its initial release thanks to modders~\cite{skyrimpopular}. 
Thus, game modding can help alleviate the pressure on game developers of having to provide new game content. In some extreme cases, such new game content can even lead to a new standalone game. For example, the \textit{Defense of the ancients (Dota)} game series\footnote{\url{https://www.dota2.com/play/}} originated as a mod for the \textit{Warcraft 3} game,\footnote{\url{http://us.blizzard.com/en-us/games/war3/}} and then became a standalone game due to its popularity. The \textit{Dota 2} game still has an average of approximately 450k players every day, more than 5 years after its initial release and more than 16 years after the release of the \textit{Warcraft 3} game~\cite{dota_players}.

In this paper, we study 9,521 mods of the 20 most-modded games on the Nexus Mods distribution platform,\footnote{\url{https://www.nexusmods.com/}} one of the largest online distribution platforms for game mods. Our goal is to provide insights into the modding community of the Nexus Mods distribution platform from a quantitative perspective, and to provide researchers with a solid foundation for future exploration of game mods. In doing so, game developers can potentially reduce development time and cost due to the increased replayability of their games through mods. We compare the mods on the Nexus Mods distribution platform along three dimensions: (1)~media (e.g., visual improvements) and non-media mods (e.g., complete overhauls), (2)~mods with and without official modding support from the game developer, and (3)~mods for Bethesda and non-Bethesda games (as the Bethesda game studio is known to be very supportive towards modders). Overall, we want to study the characteristics of popular mods, and we want to investigate how these mods are maintained. 

We first conduct a preliminary study to understand the type of mods that are
created by modders. We observed that games with a better modding support from the original game developers (e.g., through an official modding tool or API) tend to have a more active modding community. An important feature of software distribution platforms (including those for game mods) is to ease the deployment of software releases. Our paper focuses on two aspects related to game mod releases: their release schedules and the post-release support. Both these aspects are well-studied by other software engineering researchers for other types of distribution platforms (such as mobile app stores~\cite{hassan_badupdates,hassan2017eu,villarroel2016release} or game distribution platforms~\cite{lin2017eag,urgentupdates}). We address the following two research questions (RQs):
 
\begin{description}

\item [\textbf{RQ1:}] \textbf{What is the release schedule of mods?} A mod has a median of two releases, which are often made closely after each other. Providing an officially-supported modding tool for a game is associated with mods being released faster, in particular when that tool is available for several games within (or even across) the game franchise.

\item [\textbf{RQ2:}] \textbf{How well is the post-release support of mods?} In general, bug reports for mods are of low quality. A very small portion (approximately 2 to 4\%) of the bug reports contain a code sample or a stack trace. Hence, it is not surprising that 67\% of the bug reports still have the `New issue' status. Most bug reports are discussed only by the bug reporter and mod developer. However, we came across many informal bug reports outside of the official bug reporting tool. Hence, users are willing to submit bug reports, but they fail to do so in a systematic manner, which leads to low quality bug reports.

\end{description}

The remainder of the paper is outlined as follows. Section~\ref{sec:bg} gives background information about the Nexus Mods distribution platform. Section~\ref{sec:related} gives an overview of related work. Section~\ref{sec:method} discusses our methodology. Section~\ref{sec:prelimstudy} discuss our preliminary study of the characteristics of mods. Sections~\ref{sec:rq1} and~\ref{sec:rq2} discuss the results of our empirical study. Section~\ref{sec:implications} discusses the implications of our study. Section~\ref{sec:challenges} discusses the research challenges that we identified for future researchers of game mods. Section~\ref{sec:threats} outlines threats to the validity of our findings. Section~\ref{sec:conclusion} concludes our study.

\section{The Nexus Mods Distribution Platform}
\label{sec:bg}

This section gives a brief overview of the Nexus Mods distribution platform. The Nexus Mods distribution platform is one of the largest distribution platforms for game mods with over 230,000 downloadable mods~\cite{populargamemods}. Table~\ref{tab:mod_platform_overview} shows a comparison of the number of mods on other popular distribution platforms for mods.

All mods on the Nexus Mods distribution platform are free to download. Each mod has a publisher and a creator: the \emph{mod publisher} is the user who uploaded or distributed the mod and the \emph{mod creator} is a self-defined name by the mod publisher. Hence, the creator data does not necessarily reflect the actual creator of the mod. Based on our observation, the mod creator data is noisy since it is self-defined; e.g., the mod creator name ``me" is frequently specified as the mod creator. Therefore, we refer to the mod publisher as the modder in the remainder of this paper. In some cases, the mod was published by the original game developer. For example, the CD Projekt Red game studio published their MODKit for the Witcher 3 game through the Nexus Mods platform~\cite{modkit_nexus}.

The Nexus Mods distribution platform has a community of over 14 million registered users. Mod developers can distribute their mods through Nexus Mods, but also upload new releases of their mods (along with release notes). 
In addition, the Nexus Mods distribution platform offers its own simple bug reporting system,\footnote{\url{https://www.nexusmods.com/news/12474}} which allows users who downloaded a mod to report a bug for that mod. Mod developers have the option to enable or disable the bug reporting system on their mod’s webpage. Each user has the option to add a title and a description, and the choice of making the report private. In addition, there are four bug report statuses on the Nexus Mods distribution platform that we consider to be open (i.e., \emph{new issue}, \emph{being looked at}, \emph{known issue}, and \emph{needs more info}), whereas we consider the other statuses (i.e., \emph{fixed}, \emph{not a bug}, \emph{won't fix}, and \emph{duplicate}) as resolved statuses.

Each mod has a dedicated mod page, which displays metadata such as the mod title, mod publisher, mod category, the number of endorsements, the number of unique downloads, and the files of the mod. In addition, the mod page can include a detailed description, images or videos, forum posts, mod statistics and bug reports. Endorsements\footnote{\url{https://help.nexusmods.com/article/45-what-are-file-endorsements}} are the Nexus Mods distribution platform's version of a rating and represents the user's appreciation of the mod. An endorsement can be considered an `upvote' of the associated mod. Nexus Mods does not provide a mechanism for giving a negative endorsement (other than removing a prior endorsement of the mod). Endorsements help other users find useful or enjoyable mods, but does not necessarily reflect the quality of the mod. In order to provide an endorsement, the mod must be downloaded first. 

\begin{table}[!t]
\caption{Online Mod Distribution Platforms Overview}
\centering
\begin{tabular}{lr}
\toprule
Mod Distribution Platforms & \# of Mods \\
\midrule
Nexus Mods& 230,581 \\
Game Modding$^1$& 71,262 \\
Moddb$^2$& 18,293 \\
Mods Online$^3$& 4,766 \\
\bottomrule
\multicolumn{2}{l}{\small{$^1$ \url{http://www.gamemodding.net/}}} \\
\multicolumn{2}{l}{\small{$^2$ \url{https://www.moddb.com/}}} \\
\multicolumn{2}{l}{\small{$^3$ \url{http://modsonline.com/}}} \\
\end{tabular}
\label{tab:mod_platform_overview}
\end{table}

\section{Related Work}
\label{sec:related}

This section discusses work that is related to our study. We discuss work on (1)~game modding, (2)~games and software engineering and (3)~mining online distribution platforms.

\emph{Game Modding:}
The majority of prior work on game modding focused on the cultural, social and economical aspects of modding. For example, several studies focused on the types of mods that exist and the motivations of developers and players for modding and using mods~\cite{historyofmodding, playermotivations, modthesis, ofmodandmodders, moddermotivation}. In addition, several studies investigated the relation and mutual benefits between a game and its mod developers~\cite{profitingfrominnovativeuser, industriallogic, digitalconsumernetworks, whowroteeldderscrolls, modintermedialityculture}.  Poretski and Arazy~\cite{placingvalueoncommunitycocreations} conducted an empirical study on 64 games from the Nexus Mods distribution platform and found value in mods with respect to an increase in sales of the original game. 

There are relatively few studies on game modding from a software engineering perspective. Dey et al.~\cite{populargamemods} studied the mod-popularity of features in the 6 most-modded games on the Nexus Mods distribution platform by looking at the number of unique downloads and associated tags with each mod. They found that untagged mods are the least popular.
Scacchi~\cite{moddingsystem, modopensource,scacchi_firstmonday} gives an overview of the types of mods and the modding community, and describes modding as an open source approach to extending closed source software. 
We are the first to conduct a large-scale study of the release schedule and bug reports of game mods.

\emph{Games and Software Engineering:} Several studies focused on games and software engineering. Several studies mined data from game projects to study software engineering aspects. For example, Ahmed et al.~\cite{ahmed2017open} and Pascarella et al.~\cite{gamevsnongame} analyzed open source game projects. Graham and Roberts~\cite{graham2006toward} and K\"ohler et al.~\cite{kohler2012feedback} studied the development of their own game, whereas Guana et al.~\cite{guana2015building} studied the development of their own game engine. B{\'e}cares et al.~\cite{becares2017approach} studied the gameplay of a small game called Time and Space. In addition, several studies~\cite{lewis2010went,lin2019identifying} investigated what can be learned from bug videos.

Other studies investigated software engineering aspects by mining data from online game distribution platforms. Lin et al. studied the impact of the early access release model~\cite{lin2017eag}, emergency updates~\cite{urgentupdates} and reviews~\cite{lin2018gr} of games on the Steam online distribution platform. Cheung et al.~\cite{cheung2014first} analyzed over 200 reviews for Xbox 360 games. 

A few prior studies mined data from game literature (such as articles and magazines). For example, several studies~\cite{lewis2011whats,asurveyonactualsoftwareengineeringprocesses,Washburn2016}  mined post-mortems of games for insights about what went wrong and what went right during the game development. Scacchi and Cooper~\cite{scacchi2015research} studied literature on computer games and software engineering. In addition, Ampatzoglou and Stamelos~\cite{ampatzoglou2010software} systematically reviewed existing literature on software engineering for games.

A few prior studies studied game development through interviews~\cite{strivingforbalance, videogamedevelopmentdifferentfromsoftwaredev} and surveys~\cite{whatconcernsgamedevelopers,koleva2015front}.

The above studies focused on several aspects of games and software engineering, such as software engineering practices applied in game development~\cite{gamevsnongame,Washburn2016,whatconcernsgamedevelopers,graham2006toward,guana2015building,asurveyonactualsoftwareengineeringprocesses}, limitations of game development~\cite{koleva2015front,videogamedevelopmentdifferentfromsoftwaredev,ahmed2017open}, game testing~\cite{becares2017approach,kohler2012feedback}, empirical insights on game development for game developers~\cite{lewis2010went,urgentupdates,lin2017eag,lin2018gr,lin2019identifying}, user views of games~\cite{cheung2014first,strivingforbalance}, and potential software engineering research opportunities in games~\cite{ampatzoglou2010software,lewis2011whats,scacchi2015research}. 

The above papers on games and software engineering all focus on the development of the original game. In contrast, our paper focuses on the development of game mods. Since the development teams of game mods and original games are likely considerably different (e.g., most original games are developed by large companies, while mods are often developed by individuals), our paper is the first work to explore this very important complementary community given that game mods are likely a very different type of software than games.

\emph{Mining Online Distribution Platforms:}
Studies on mining large game distribution platforms focused mostly on the Steam distribution platform. The work of Lin et al.~\cite{lin2017eag, lin2018gr, urgentupdates, lin2019identifying} (described above) mined the Steam platform from a software engineering point of view. Sifa et al.~\cite{theplaytimeprinciple} analyzed the playtime of 6 million users on the Steam platform and observed that there were fundamental principles regarding playtime. 
Blackburn et al.~\cite{cheatinginonlinegamesasocialnetwork} studied cheaters in the Steam gaming community and observed that the player's network of friends impacted the likelihood of a player becoming a cheater.

The large body of prior work on mobile app analysis focused mostly on the Google Play store and Apple's App Store. As a complete overview of prior work in this field is outside of the scope of this paper, we refer to the survey of Martin et al.~\cite{martin2017survey} for such an overview. 
Several studies focused on release schedules of mobile apps. Hassan et al.~\cite{hassan2017eu} studied the frequency of emergency updates, and found eight patterns of emergency releases. In addition, Hassan et al.~\cite{hassan_badupdates} showed that it is important to study mobile app reviews at the release-level rather than at the update-level. 
McIlroy et al.~\cite{frequentlyupdatedapps} studied the update frequency in 10,713 mobile apps that were mined from the Google Play store and observed that only a small subset of the apps were updated more than once a week, and that 45\% of the updates did not provide a rationale for the update.

\section{Methodology}
\label{sec:method}

\begin{figure*}[!t]
\center
    \includegraphics[width=1\textwidth]{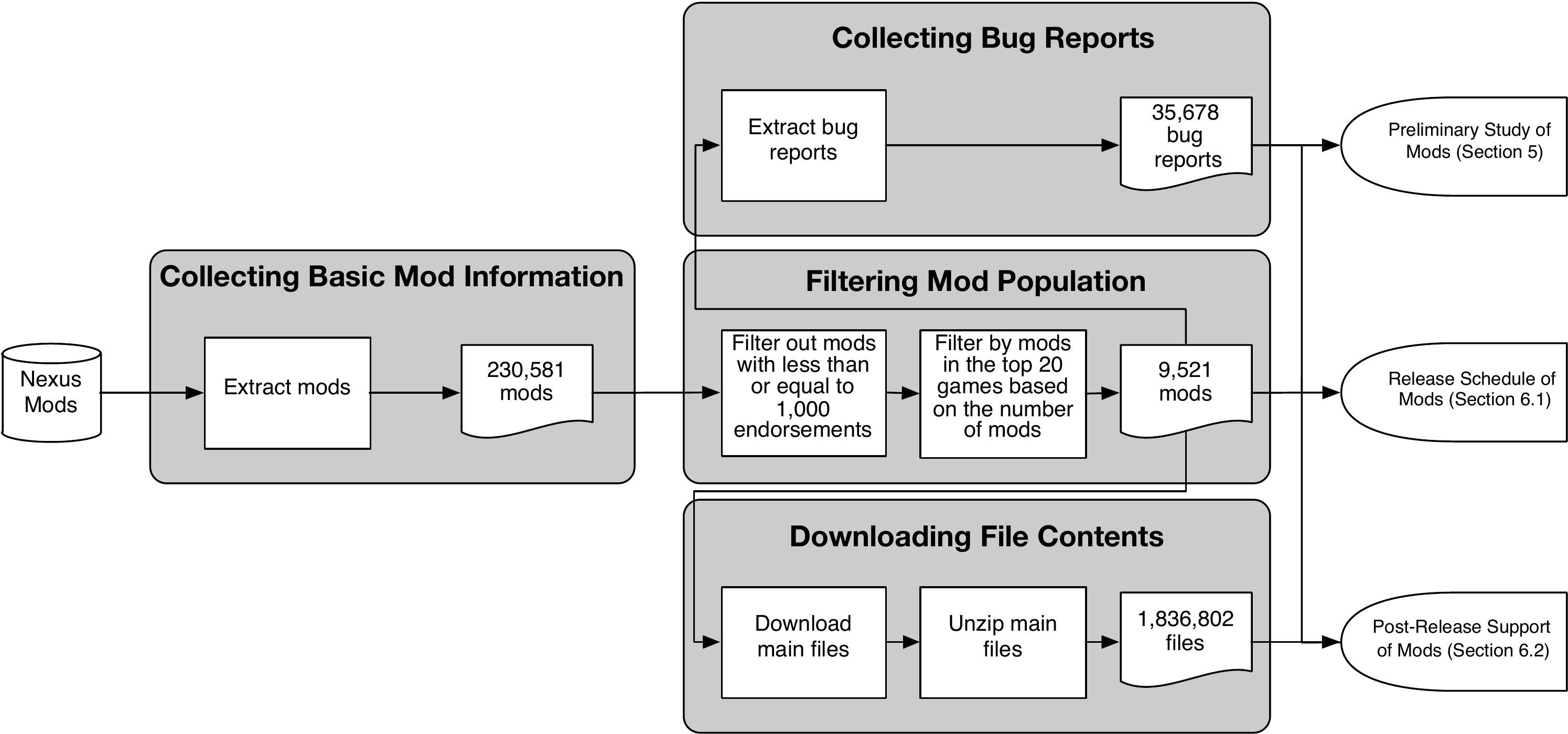}
  \caption{An overview of our methodology.}
  \label{fig:methodology_overview}
\end{figure*}

This section discusses the methodology of our empirical study of game mods. Figure~\ref{fig:methodology_overview} presents the steps of our methodology. In addition, our dataset is available in the supplementary material of our paper~\cite{supp_material}.

\subsection{Collecting Basic Mod Information}

We extracted 230,581 mods from the Nexus Mods distribution platform using a customized crawler on January~29, 2018. The mod metadata that we used in our study consists of the mod title, mod category, mod file size, mod files, total number of endorsements, total number of unique downloads, and the total number of bug reports. 

\subsection{Filtering Unpopular Mods}

\begin{table} [!t]
	\centering
	\caption{An overview of the studied mods.} 
	\label{fig:mod_dataset_overview}
	\begin{tabular} {lr}
	\toprule
	\bfseries Total number of mods & 230,581 \\
	\bfseries Number of studied mods & 9,521 \\
	\addlinespace
	\bfseries Total number of games & 502 \\
	\bfseries Number of studied games & 20 \\
	\addlinespace
	\bfseries Number of studied mods with files & 9,370 \\
	\bfseries Number of studied files & 1,836,802 \\
	\addlinespace
	\bfseries Number of studied bug reports& 35,678 \\
	\bfseries Number of studied bug comments& 109,133 \\
	\bottomrule
	\end{tabular}
\end{table}
 
We observed that there are many mods on the Nexus Mods platform which do not work or are not used by anyone. As such mods are not representative of the mods that are potentially helpful to game developers, we studied only a subset of the collected mods. We had the following selection criteria for the studied mods:

\begin{itemize}
\item \textbf{Quality}: The mod is of reasonable quality.
\item \textbf{Popularity}: The mod has a large community. Hence, such a mod would reflect what gamers would like to see in a game.

\end{itemize}

The first criteria is satisfied by selecting mods that do not only have a large community, but are also widely appreciated (i.e., at least 1,000 endorsements). The second criteria is satisfied by selecting only mods from the top 20 most-modded games on Nexus Mods, which left us with a total of 9,521 mods. This criterion ensures that our findings are not biased towards unpopular games with a small number of mods. Table~\ref{fig:mod_dataset_overview} gives an overview of the studied mods. Table~\ref{tab:top_20_game_overview} gives an overview of the studied games and their number of mods.

\subsection{Collecting Bug Reports}
As explained in Section~\ref{sec:bg}, the bug reporting system of the Nexus Mods distribution platform only supports bug reports. As Table~\ref{tab:bug_status_definition} shows there exists a bug status ``Not a bug'' in the Nexus Mods bug reporting system. We assume that mod developers leverage this bug status to filter out reports that do not discuss a bug but for example, request a feature. Hence, we assume that all bug reports without this status discuss actual bugs. We collected the bug report metadata of each studied mod. In total, we collected 35,678 bug reports for 2,160 out of the 9,521 studied mods. At the time of crawling, the publishers of the remaining 7,361 mods explicitly disabled the bug reporting functionality. In addition, we collected the comments on each studied bug report. The total number of collected bug comments is 109,133.

\begin{table}[!t]
\caption{Our studied games (sorted by the number of mods).}
\centering
\resizebox{\textwidth}{!}{

\begin{tabular}{lrrll}
\toprule
\multicolumn{5}{l}{} \\
Game Title & Year of Release & \# of Mods & Game Genre\textsuperscript{1} & Game Developer \\
\midrule
The Elder Scrolls: Skyrim\textsuperscript{2} & 2011 & 4,793 & Action role-playing & Bethesda Game Studios \\  
Fallout 4 & 2015 & 1,675 & Action role-playing & Bethesda Game Studios \\  
Fallout: New Vegas & 2010 & 974 & Action role-playing & Bethesda Game Studios \\  
The Elder Scrolls: Skyrim Special Edition\textsuperscript{2} & 2016 & 518 & Action role-playing & Bethesda Game Studios \\  
The Elder Scrolls: Oblivion\textsuperscript{2} & 2006 & 442 & Action role-playing & Bethesda Game Studios \\ 
\addlinespace 
Fallout 3 & 2008 & 387 & Action role-playing & Bethesda Game Studios \\  
Dragon Age & 2009 & 224 & Action role-playing & BioWare \\  
The Witcher 3 & 2015 & 140 & Action role-playing & CD Projekt Red \\  
Dragon Age: Inquisition & 2014 & 76 & Action role-playing & BioWare \\  
The Elder Scrolls: Morrowind\textsuperscript{2} & 2002 & 65 & Action role-playing & Bethesda Game Studios \\  
\addlinespace
Dragon Age 2 & 2011 & 35 & Action role-playing & BioWare \\  
Dark Souls & 2011 & 33 & Action role-playing & FromSoftware \\  
The Witcher 2 & 2011 & 27 & Action role-playing & CD Projekt Red \\  
Stardew Valley & 2016 & 23 & Simulation & ConcernedApe\\  
Mass Effect 3 & 2012 & 23 & Action role-playing & BioWare\\ 

\addlinespace 
Mount \& Blade: Warband & 2010 & 21 & Action role-playing & TaleWorlds Entertainment\\  
XCOM 2 & 2016 & 20 & Turn-based tactics & Firaxis Games\\  
State of Decay & 2013 & 16 & Action-adventure & Undead Labs\\
The Witcher & 2007 & 15 & Action role-playing & CD Projekt Red\\  
XCOM: Enemy Unknown & 2012 & 14 & Turn-based tactics & Firaxis Games\\
\addlinespace
\bottomrule
\end{tabular}}
\label{tab:top_20_game_overview}
\begin{minipage}[t]{11.4cm}

	\begin{tablenotes}
 \small
	\item[1]The game genre was obtained from Wikipedia
	\item[2]Skyrim/Skyrim SE/Oblivion/Morrowind hereafter
\end{tablenotes}
\end{minipage}
\end{table}

\subsection{Determining Whether a Game has Official Modding Support}
We determined if the original game developers provided modding support for a game by manually searching various sources, such as the game's official website, to find whether the original game developers released a modding tool. We identified an official modding tool for 65\% of the studied games.

\subsection{Downloading File Contents and Identifying Release Dates}

The Nexus Mods platform hosts files in several categories for each mod. We collected files from the following categories:

\begin{itemize}
\item \textbf{Main files:} These files are the base files of the latest version of a mod and usually come as a compressed archive. We used these files to study the file types that are modified or added by a mod. After downloading and decompressing the main files, we used the \code{File}\footnote{\url{https://linux.die.net/man/1/file}} program to automatically classify the file type of each decompressed file. Note that it is not sufficient to simply do the file type classification based on the file extension, as we observed that different games might use the same extension for different file types. We automatically classified the file type of 1,836,802 files. 
\item \textbf{Update files and old files:} These files are older versions of the main files. We used the release dates of these files to identify the release schedule of a mod. 
\end{itemize}

\subsection{The Studied Dimensions of Mods}

\begin{figure*}[!t]
\center
    \includegraphics[width=0.65\textwidth]{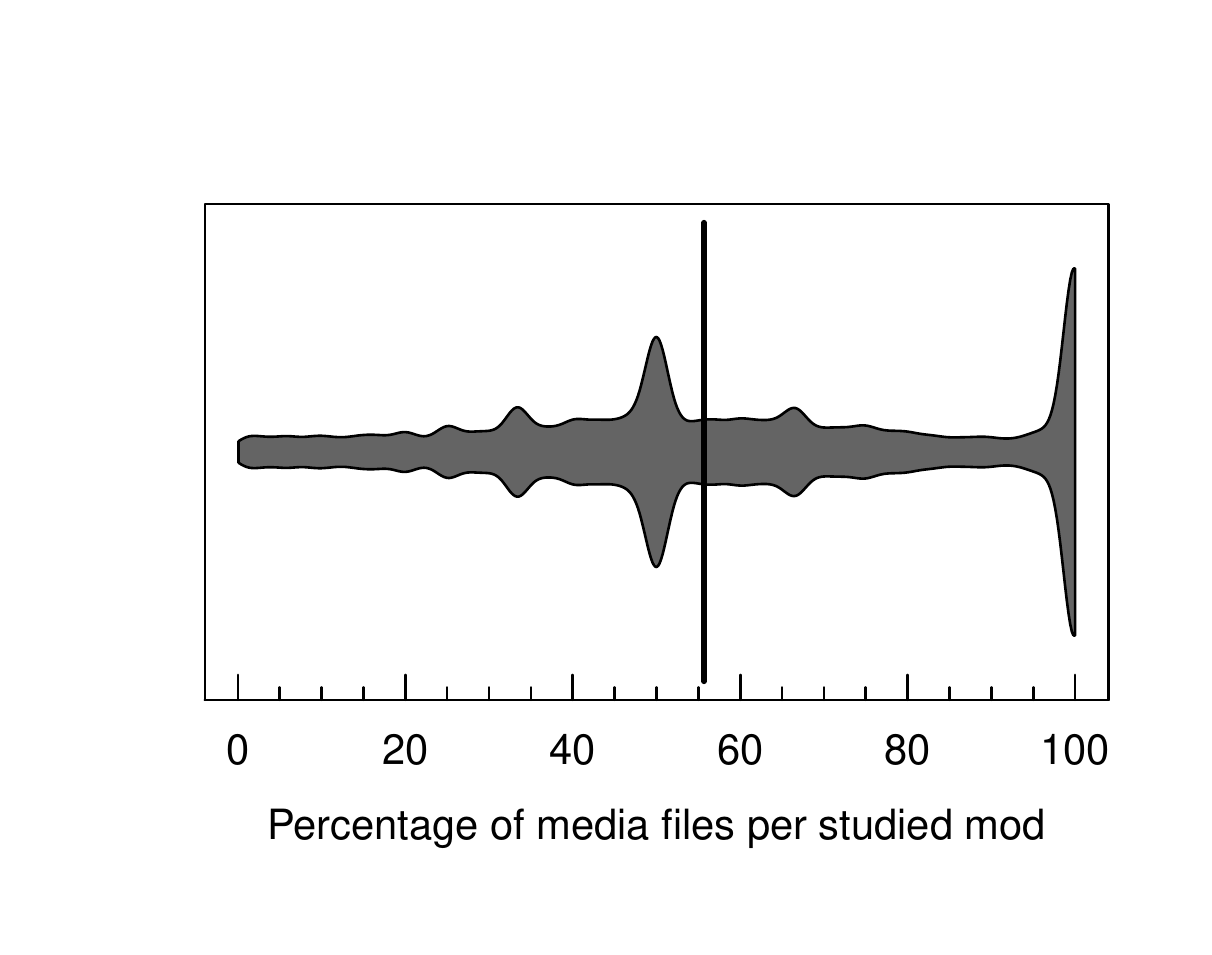}
  \caption{The percentage of media files in each studied mod. The black vertical line shows the median value.}
  \label{fig:media_file_percentage}
\end{figure*}

We study game mods along the following three dimensions:

\begin{enumerate}
\item \textbf{Games with and without official modding support.} While some games try to prevent modding (e.g., multiplayer games to avoid cheating), other games provide official modding support. Games with official modding support (e.g., through a tool or an API that is provided by the game's developer) may make it easier to develop and maintain mods, which could promote an active modding community. 
Table~\ref{tab:top_20_modding_tools} gives an overview of the official modding tools of the studied games. We studied 9,315 mods of games with official modding support and 206 mods of games without official modding support.

\item \textbf{Media and non-media mods.} Media mods (e.g., mods that change the look-and-feel of a game) require less programming effort than non-media mods (e.g., mods that change the gameplay of a game)~\cite{mediavsnonmedia}. Our hypothesis is that non-media mods are harder to create and maintain. The following explains our approach to distinguish between media and non-media mods:

\begin{enumerate}
\item We collected 140 distinct file types from the studied mods. We removed file types that took up less than 1\% of the total files to remove the bias of infrequently used file types, leaving 37 distinct file types. We calculated the percentage of files of each file type within each studied game. 
\item The first and third author independently categorized the 37 file types into high-level file categories. Table~\ref{tab:file_types} shows the file types that reside in each file category. The process of grouping the file types was straight-forward because each file type had a distinct functionality. However, we did merge the audio and video file types together into a multimedia file category. After consolidating their categorization and resolving the conflicts, we ended up with 8 high-level file categories (archive, font, binary, source code, text/documentation, multimedia, image, and game data).
\begin{table}[!t]
\caption{An overview of the official modding tools of the studied games, sorted by the total number of mods.}
\centering
\resizebox{\textwidth}{!}{
\begin{tabular}{llll}
\toprule
Game Title & Official Tool & Official URL \\
\midrule
Skyrim & Creation Kit & \url{https://www.creationkit.com/} \\  
Fallout 4 & Creation Kit & \url{https://www.creationkit.com/} \\ 
Fallout: New Vegas & GECK & \url{https://geck.bethsoft.com/}  \\  
Skyrim Special Edition & Creation Kit & \url{https://www.creationkit.com/}  \\    
Oblivion & Construction Set & \url{https://cs.elderscrolls.com} \\ 
\addlinespace 
Fallout 3 & GECK & \url{https://geck.bethsoft.com/} \\  
Dragon Age & Toolset & \url{http://www.datoolset.net/}  \\  
The Witcher 3 & MODKit & \url{https://www.nexusmods.com/witcher3/mods/3173}  \\  
Dragon Age: Inquisition & - & - & -\\  
Morrowind & Construction Set & \url{https://cs.elderscrolls.com}  \\  
\addlinespace
Dragon Age 2 & Toolset & \url{http://www.datoolset.net/} \\  
Dark Souls & - & -  \\  
The Witcher 2 & REDKit & \url{https://redkitwiki.cdprojektred.com/REDkit} \\  
Stardew Valley & - & - \\  
Mass Effect 3 & - & -  \\ 

\addlinespace 
Mount \& Blade: Warband & - & - \\  
XCOM 2 & Modbuddy & -   \\
State of Decay & - & - \\
The Witcher & D'jinni & - \\  
XCOM: Enemy Unknown & - & -\\
\addlinespace
\bottomrule
\end{tabular}}
\label{tab:top_20_modding_tools}
\end{table}
\item We then categorized each file in the image, font, archive, text/documentation, and multimedia file categories as a media file. We categorized each file in the binary, source code, or game data file categories as a non-media file.
\item Figure~\ref{fig:media_file_percentage} shows the distribution of the percentage of media files in the studied mods. While there appears to be a natural cutoff at approximately 95\% media files, this cutoff would introduce bias for mods with a small number of files. We calculated that 57\% of the studied mods have atmost 20 files. Hence, we used thresholds that are based on the number of files to accurately determine if a mod is a media or non-media mod. If the percentage of media files in a mod is larger than the threshold for the number of files in the mod, the mod is categorized as a media mod. Otherwise, the mod is categorized as a non-media mod. We used the following thresholds:
\begin{equation*}
 \text{Media mod thresholds}=\begin{cases}
   >95\%, & \text{if \# of files $\geq 20$}.\\
   >90\%, & \text{if $20 > \#$ of files $\geq 10$}.\\
   >\Big(\dfrac{\text{\# of files} - 1}{\text{\# of files}}\Big) & \text{if \# of files $< 10$}.\\
 \end{cases}
\end{equation*}
\end{enumerate}

\begin{table}[!hbtp]
\caption{An overview of the file categories and the included file types (sorted alphabetically).}
\centering

\begin{tabular}{ll}
\toprule
\textbf{File Category} & \textbf{File Types}\\
\midrule
Archive & RAR archive data \\
\addlinespace
\begin{tabular}[l]{@{}l@{}}Binary\\\\\\\\\\\\\\\end{tabular} & 
\begin{tabular}[l]{@{}l@{}}PE32 executable (console) Intel 80386\\
PE32 executable (DLL) (console) Intel 80386\\
PE32 executable (DLL) (console) Intel 80386 Mono/.Net assembly\\
PE32 executable (DLL) (GUI) Intel 80386\\
PE32 executable (DLL) (GUI) Intel 80386 Mono/.Net assembly\\
PE32 executable (GUI) Intel 80386\\
PE32 executable (GUI) Intel 80386 Mono/.Net assembly
\end{tabular}\\
\addlinespace
Font & TrueType font data \\
\addlinespace
\begin{tabular}[l]{@{}l@{}}Game Data\\\\\\\\\\\\\end{tabular} & \begin{tabular}[l]{@{}l@{}}Data\\
Gamebryo game engine file\\
MSVC program database ver 7.00\\
NetImmerse game engine file\\
Unreal Engine Package\\
XML 1.0 document text

\end{tabular} \\
\addlinespace
\begin{tabular}[l]{@{}l@{}}Image\\\\\\\\\\\\\end{tabular} & \begin{tabular}[l]{@{}l@{}}
JPEG image data\\
Microsoft DirectDraw Surface (DDS)\\
PC bitmap\\
PNG image data\\
Targa image data - RGB 8 x 8 x 24 Targa image data - RGB 8 x 8 x 24\\
Targa image data - RGBA - RLE 32 x 32 x 32 - 8-bit alpha
\end{tabular}\\
\addlinespace
\begin{tabular}[l]{@{}l@{}}Multimedia\\\\\\\\\\\end{tabular} & \begin{tabular}[l]{@{}l@{}}
Audio file with ID3 version 2.3.0\\
Bink Video rev.i\\
MPEG ADTS\\
Ogg data\\
RIFF (little-endian) data
\end{tabular}\\
\addlinespace
\begin{tabular}[l]{@{}l@{}}Source Code\\\\\\\\\end{tabular} & \begin{tabular}[l]{@{}l@{}}
Algol 68 source text\\
C program text\\
C++ source text\\
Python script text executable
\end{tabular}\\
\addlinespace
\begin{tabular}[l]{@{}l@{}}Text/Documentation\\\\\\\\\\\\\\\end{tabular} & \begin{tabular}[l]{@{}l@{}}
ASCII text\\
HTML document text\\
ISO-8859 text\\
Little-endian UTF-16 Unicode text\\
Unified diff output text\\
UTF-8 Unicode (with BOM) text\\
UTF-8 Unicode text
\end{tabular}\\
\addlinespace
\bottomrule
\end{tabular}
\label{tab:file_types}

\end{table}

We studied 1,348 media mods and 8,145 non-media mods.

\item \textbf{Mods of Bethesda and non-Bethesda games.} 
Table~\ref{tab:top_20_game_overview} shows that 7 of the 20 studied games were produced by Bethesda Game Studios (hereafter Bethesda). Comparing mods of Bethesda games with mods of non-Bethesda games can yield interesting insights into why Bethesda games have such active modding communities.
We studied 8,829 mods of Bethesda games and 664 mods of non-Bethesda games.

\end{enumerate}

Throughout our study, we used the Wilcoxon test to assess whether the two groups within each of the abovementioned dimensions differ significantly. The Wilcoxon rank-sum test is an unpaired, non-parametric statistical test, where the null hypothesis is that two distributions are identical, while the Wilcoxon signed-rank test is a paired non-parametric statistical test~\cite{wilcoxon1945individual}. If the p-value of the applied Wilcoxon test is less than 0.05, then the two distributions are significantly different under a significance threshold of $\alpha$ = 0.05. To calculate the magnitude of the difference between the two distributions, we compute Cliff's delta \textit{d}~\cite{long2003ordinal} effect size. We use the following thresholds for \textit{d}~\cite{romano2006exploring} :
\begin{equation*}
 \text{Effect size}=\begin{cases}
   \textit{negligible(N)}, & \text{if $|d| \leq 0.147$}.\\
   \textit{small(S)}, & \text{if $0.147 < |d| \leq 0.33$}.\\
   \textit{medium(M)}, &\text{if $0.33 < |d| \leq 0.474$}.\\
   \textit{large(L)}, & \text{if $0.474 < |d| \leq 1$}. \\
 \end{cases}
\end{equation*}

In the next sections, we discuss the results of our study. First we discuss our preliminary study.

\section{Preliminary Study of Mods from the Nexus Mods Distribution Platform}
\label{sec:prelimstudy}
Prior research used the total number of unique downloads of a mod as a proxy for its popularity~\cite{populargamemods}. However, a download does not necessarily mean that a gamer enjoyed or played the mod. An endorsement is a better indication that a gamer has played and appreciated the mod. Therefore, we studied the characteristics of highly endorsed mods to gain a better understanding of the mods that are appreciated by the modding community. 

\begin{table}[!t]
\caption{The definitions of the mod categories (sorted alphabetically).}
\centering
\resizebox{\textwidth}{!}{
\begin{tabular}{llr}
\toprule
\multicolumn{1}{c}{\textbf{Mod Category}} & 
\multicolumn{1}{c}{\textbf{Definition}} & 
\multicolumn{1}{c}{\textbf{\# of studied mods}}  
\\
\midrule
Appearance            & Alters the visual features, such as the body, face, & 283\\ 
& and hair of the character & \\
Audio                 & Alters the game audio & 138\\
Bug Fixes             & Fixes a bug in the game & 240\\
Character             & Alters playable and non-playable & 873\\ & characters in the game  & \\
Cheats                & Alters the gameplay in an illegal or unjust manner & 202\\
Game Items &  Adds or alters objects that a player can collect & 2,250\\ & or use in the game & \\
Game Visuals          & Alters the observable experience of the game environment & 1,749\\
Gameplay              & Alters the player experience and interaction & 1,498 \\ 
& within the domain of the game rules & \\
Overhaul              & Total conversion of certain game features or the full game & 140 \\
Performance           & Optimizes the game to improve performance & 11 \\
Save                 & Starts the player in a specific state in the game & 15 \\
Story                 & Adds to or alters the storyline of the game & 194 \\
User Interface        & Adds to or alters the interface of the game & 296\\
Utility               & Tool that helps modders & 307 \\
World Object          & Adds or alters physical objects with which primary characters can & 770 \\ 
& interact inside the game environment & \\
\bottomrule
\end{tabular}}
\label{tab:category_definition}
\end{table}

\textit{Approach:} As the number of players can differ considerably for each game, we normalized the number of endorsements by calculating the endorsement ratio and use this ratio instead of the absolute number of endorsements:

\[\text{\emph{endorsement ratio}} = \frac{\text{\emph{\# of endorsements}}}{\text{\emph{\#\ of\ unique downloads}}}\]

To identify which type of mods are highly endorsed, we leveraged the publisher-defined mod categories. As the definition of these mod categories appears not to be moderated by Nexus Mods, the definitions contain a considerable amount of noise. The first and third author independently manually created a high-level mod categorization of the mod categories to group similar ones. For example, the mod categories `armour' and `weapons' were categorized as `game items'. The categorizations were consolidated and conflicts were resolved through discussion. The cross-validation process had a Cohen's Kappa agreement of 0.93 (which is considered very high, indicating a straightforward categorization task). Table~\ref{tab:category_definition} shows the high-level mod categories along with their  definitions. 

As the mod categories are defined by the mod publishers, we also wanted to validate their accuracy. 
Therefore, we performed a sanity check in which we manually verified for a balanced sample of 95 mods that the high-level category that we assigned to each of these mods fits their actual description. 
We validated a balanced sample to avoid bias towards mods from games with a large number of mods.
Figure~\ref{fig:downsampling_overview} shows how we created the balanced sample. Below we detail each step:

\begin{figure*}[!t]
\center
    \includegraphics[width=0.9\textwidth]{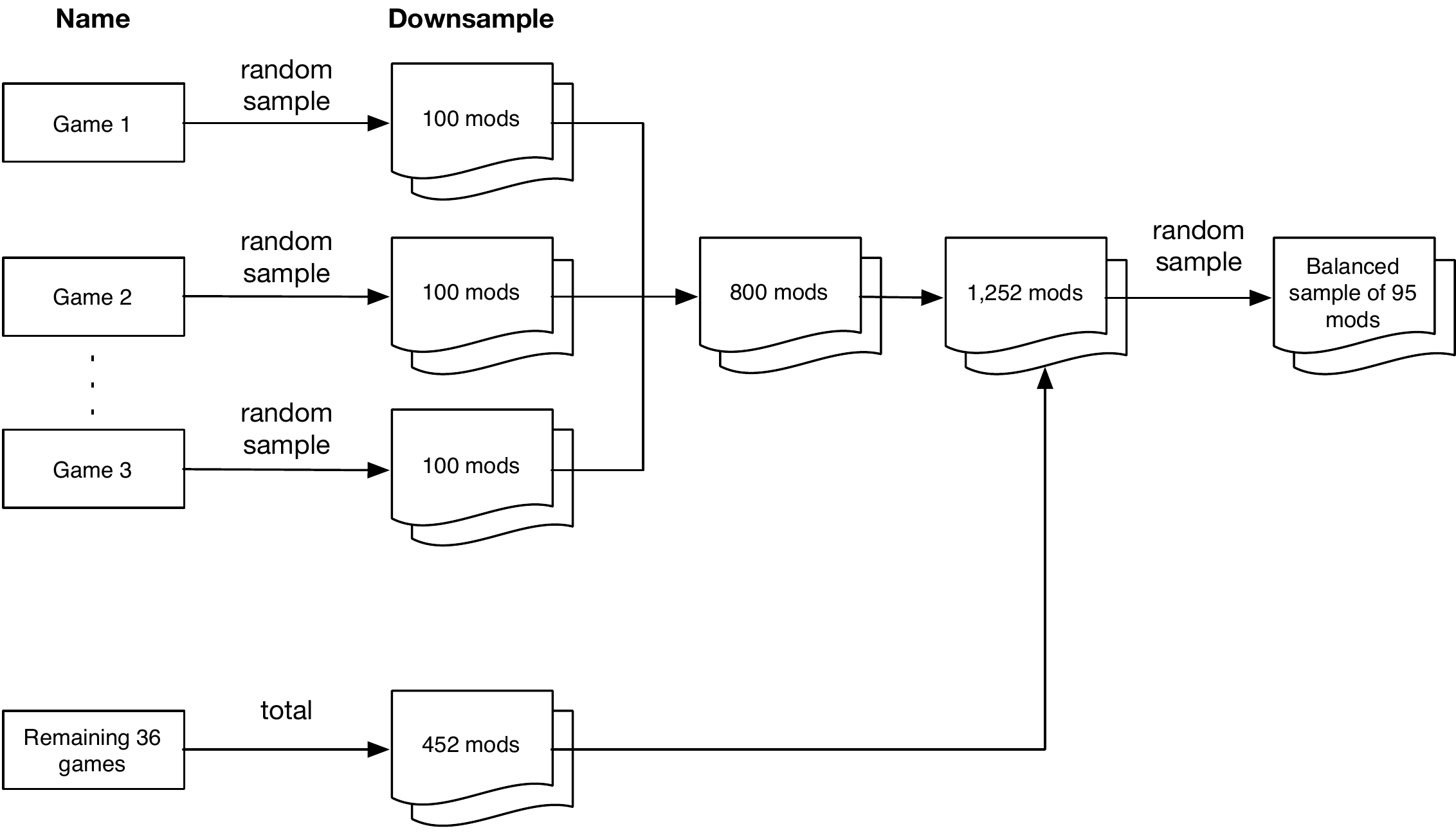}
  \caption{An overview of the creation of our balanced sample of mods.}
  \label{fig:downsampling_overview}
\end{figure*}

\begin{enumerate}
\item We select a random sample of 100 mods for each of the 8 games that have more than 100 mods (800 mods in total).
\item We select all 452 mods of the remaining 36 studied games.
\item We combine the mods selected in step 1 and 2.
\item We randomly choose a statistically representative sample of 95 mods (which yields a confidence level of 95\% with a confidence interval of 10).
\end{enumerate}	

After validating the sample of 95 mods, we found that the mod categorization of 19 of the 95 mods was incorrect or unclear. In all cases, this error or unclarity was caused by the mod publisher. Hence, our high-level mod categorization is correct for approximately 80\% of the mods. 
We did not assign a high-level category to the categories of the 19 mods that were misclassified. In total, we assigned a high-level category to 9,207 of the mods. In the rest of the paper, all analyses that were performed on the high-level categories used only these 9,207 mods.

We used the Kruskal-Wallis test to study the differences between the median endorsement ratios of the studied games across the high-level mod categories. The Kruskal-Wallis test allows us to test whether the differences between more than 2 distributions are statistically significant.

\begin{figure*}[!t]
\center
    \includegraphics[width=0.75\textwidth]{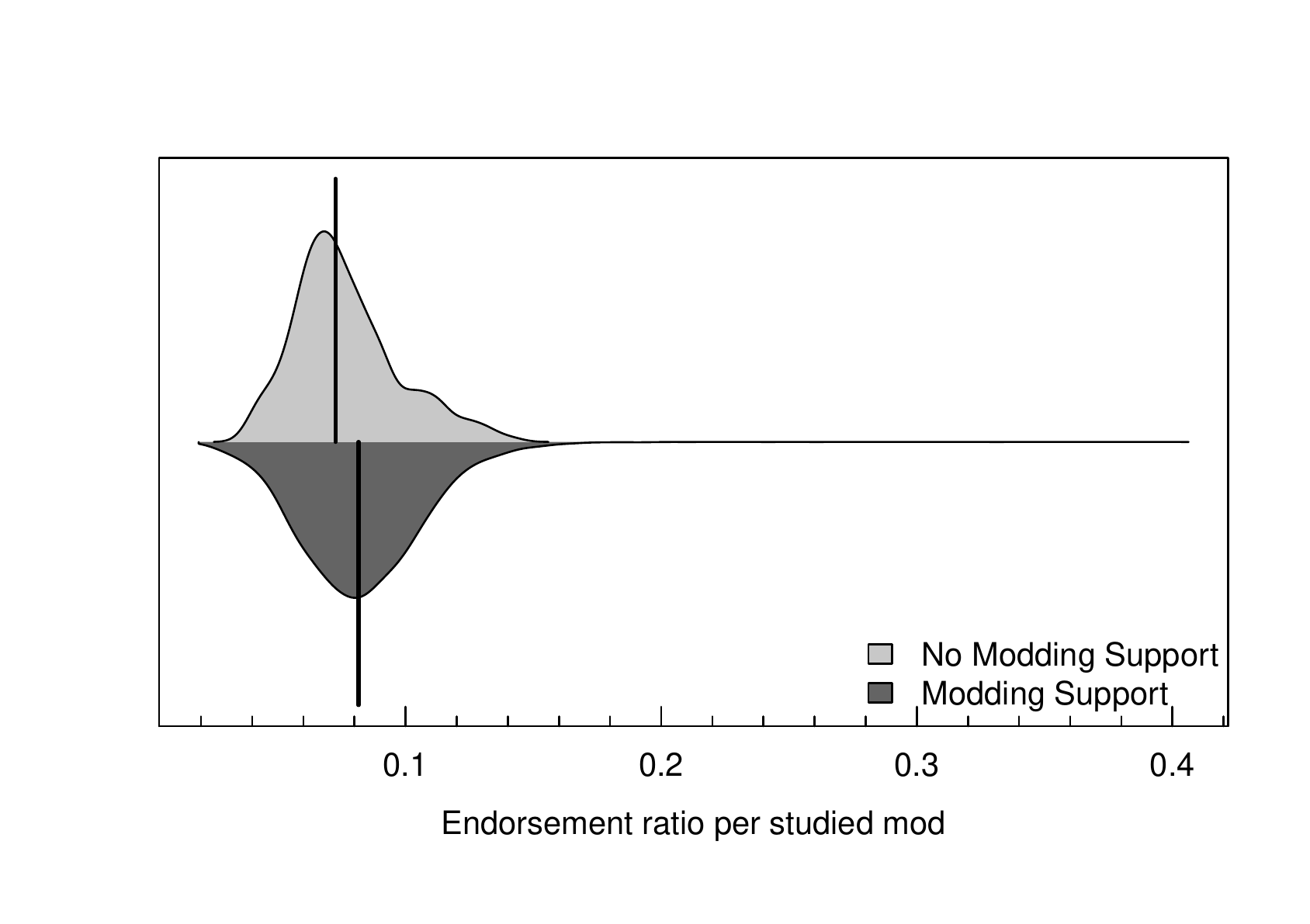}
  \caption{The distribution of the endorsement ratios per studied mod with and without official modding support. The black vertical lines are the median values.}
  \label{fig:beanplot_endorse_mod_supp_vs_no_supp}
\end{figure*}

\textit{Findings: }
\textbf{Role-Playing Games (RPGs) are the most popular games for modding.} Table~\ref{tab:top_20_game_overview} gives an overview of the studied games, sorted by the number of mods on Nexus Mods. 90\% of the studied games belong to the Role-Playing Game genre. A possible
explanation is that RPGs often allow a player to customize their game character (e.g., by earning new weapons or clothing for the game character). Hence, players of RPGs may be more interested in customizing their game character beyond the officially-supported options. Another possible explanation is that RPGs have a longer playing time compared to games of other genres, which in turn motivates players to customize their game character precisely as they wish. For example, the mod-popular \textit{Skyrim} and \textit{Fallout~4} games have an average playing time of respectively 244.9 and 186.2 hours, whereas a non-RPG game such as the \textit{State of Decay} game has an average playing time of 25.7 hours~\cite{gamelengths}. A final observation is that the vast majority of the studied games are mainly single-player games. A possible explanation is that as players can benefit more considerably from  cheating in multiplayer games~\cite{cheating_gainingadvantage}, the developers of such games take stricter measures to avoid modding.

\textbf{Mods of games with official modding support have a higher endorsement ratio than mods of games without official modding support.} Figure~\ref{fig:beanplot_endorse_mod_supp_vs_no_supp} shows that mods of games with official modding support have a median endorsement ratio of 0.08, whereas mods of games without official modding support have a median endorsement ratio of 0.07. The Wilcoxon rank-sum test shows that the two distributions are significantly different, with a small Cliff's delta effect size. While the difference between the endorsement ratios seems small, it is important to keep in mind that in general, only a small fraction of the players of a game will actually provide a rating or review of that game. For example, in our prior work~\cite{lin2017eag}, we showed that games on the Steam platform have an average review ratio of 0.01. Hence, the difference between endorsement ratios of 0.07 and 0.08 are relevant in practice. In addition, our data shows that mods on the Nexus Mods distribution platform are endorsed more often than that the original games are rated or reviewed on the Steam platform. We are unable to find a statistically significant difference under the significance threshold of $\alpha$ = 0.05 between the endorsement ratios of non-media and media mods, or between mods for non-Bethesda and Bethesda games. In addition, the Kruskal-Wallis test shows that we are unable to find a statistically significant difference in the endorsement ratio across mods from the different high-level categories.

\textbf{The most-modded games are created by Bethesda.} Table~\ref{tab:top_20_game_overview} shows that Bethesda created 7 games in the top 20 of most-modded games. A possible explanation is that Bethesda has a very strong modding support, which is well known in the modding community~\cite{bethmodpopular}. For example, Bethesda offers the Creation Kit modding tool. Table~\ref{tab:top_20_game_overview} shows that 73\% of the studied mods are for the \textit{Skyrim}, \textit{Skyrim Special Edition} and \textit{Fallout 4} games. A possible explanation for the mod-popularity of these games is the existence of the Creation Kit~\cite{modthesis}. 
A possible reason for the strong support from Bethesda for the modding community is that Bethesda uses the modding commmunity as a breeding ground for hiring new developers. For example, potential level designers were asked during the interviewing process to create a mod using the official tooling~\cite{whowroteeldderscrolls}.

\hypobox{Role-Playing Games (RPGs) are the most mod-popular games. Mods of games with official modding support are endorsed by gamers more often than those of games without such support. Officially-supported modding tools are leveraged by game developers during the hiring process of new developers or designers.}

\section{RQ1: What is the Release Schedule of Mods?}
\label{sec:rq1}

\begin{figure*}[!t]
	\center
	\includegraphics[angle=270, width=0.6\textwidth]{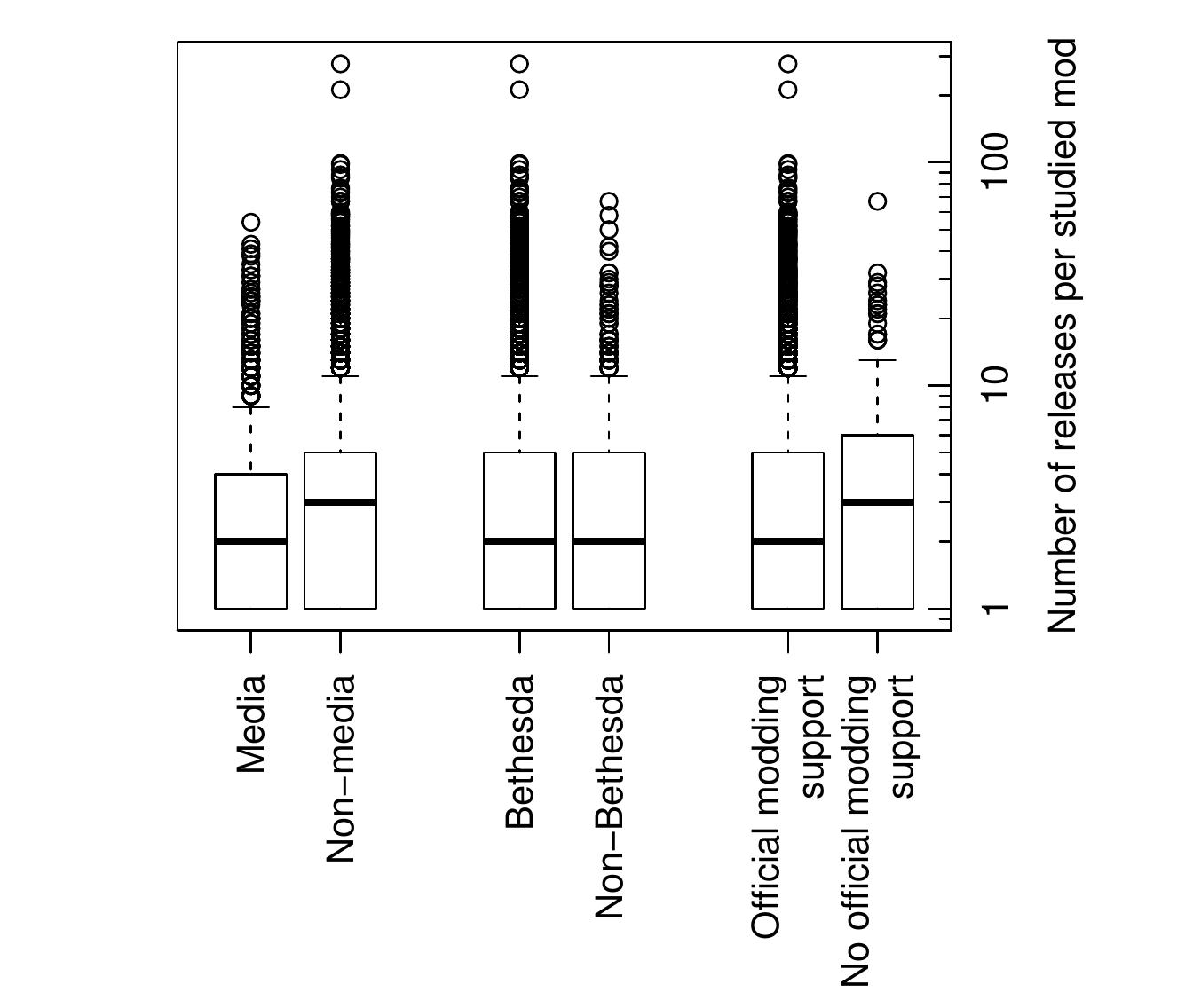}
	\caption{The distributions of the number of releases per mod across each studied dimension.}
	\label{fig:multi_num_releases}
\end{figure*}

\noindent\textit{Motivation:} 
Mods are a contribution to the original game by the modding community. 
Depending on the nature of a mod, it may require only one or several releases. For example, a mod that improves a texture in a game may require only one release. On the other hand, a mod that does a complete overhaul may require one or more follow-up releases whenever its original game releases a new version.
In this section, we study the release schedules of the mods on the Nexus Mods platform.
Assuming that a mod requires more than one release, the release schedule of a mod is an indication of the effort that a mod developer is willing to put into a mod.
In addition, we study how long it takes mod developers to release a new version of the mod after the release of a new version of the original game.

\textit{Approach:} 
To capture the release schedule of a mod, we calculated the median number of days between adjacent mod releases for each studied mod. 
To investigate the release schedule of mods after official game updates and between adjacent mod releases, we first manually collected the official release dates for each version of the studied games from several sources, such as the original game's website. 
We were able to find official release dates for 80\% of the studied games (8,804 mods).
We distinguish three release types for a mod:
\begin{itemize}
	\item \textbf{The initial mod release:} The first release of the mod after the original game release.
	\item \textbf{A back-to-back mod release:} A mod release that is adjacent in time to another mod release without a game release in between. We assume that such a release is to improve the previous mod release.
	\item \textbf{A game-triggered mod release:} A mod release that is adjacent in time to an official game release, without another mod release in between. We assume that such a release is to respond to the official game release.
\end{itemize}

In addition, we studied the time it took to release a mod for each of the games from the \textit{The Witcher}, \emph{Fallout}, \emph{XCOM}, \emph{Dragon Age} and \emph{Skyrim} game franchises. By comparing the release times for mods from games from the same franchise, we can study how official modding support (or its removal) for a game, or the official modding tool is associated with the release time of a mod. We studied one game franchise in which official modding support was added (the \emph{XCOM} franchise), one in which official modding support was removed (\emph{Dragon Age}), two franchises (\emph{The Witcher} and \emph{Fallout}) in which the officially-supported modding tool changed throughout the franchise's lifetime, and one franchise (\emph{Skyrim}) in which the officially-supported modding tool did not change. To control for the difference in age between the games within a game franchise, we focus on mods that were released within two years after the original game release. Because there were no mods in our data set for \emph{The Witcher} game that were released within the two-years time frame, we did not include this game in our study of the release time of mods within a game franchise.
	
\begin{figure*}[!t]
\center
    \includegraphics[width=0.75\textwidth]{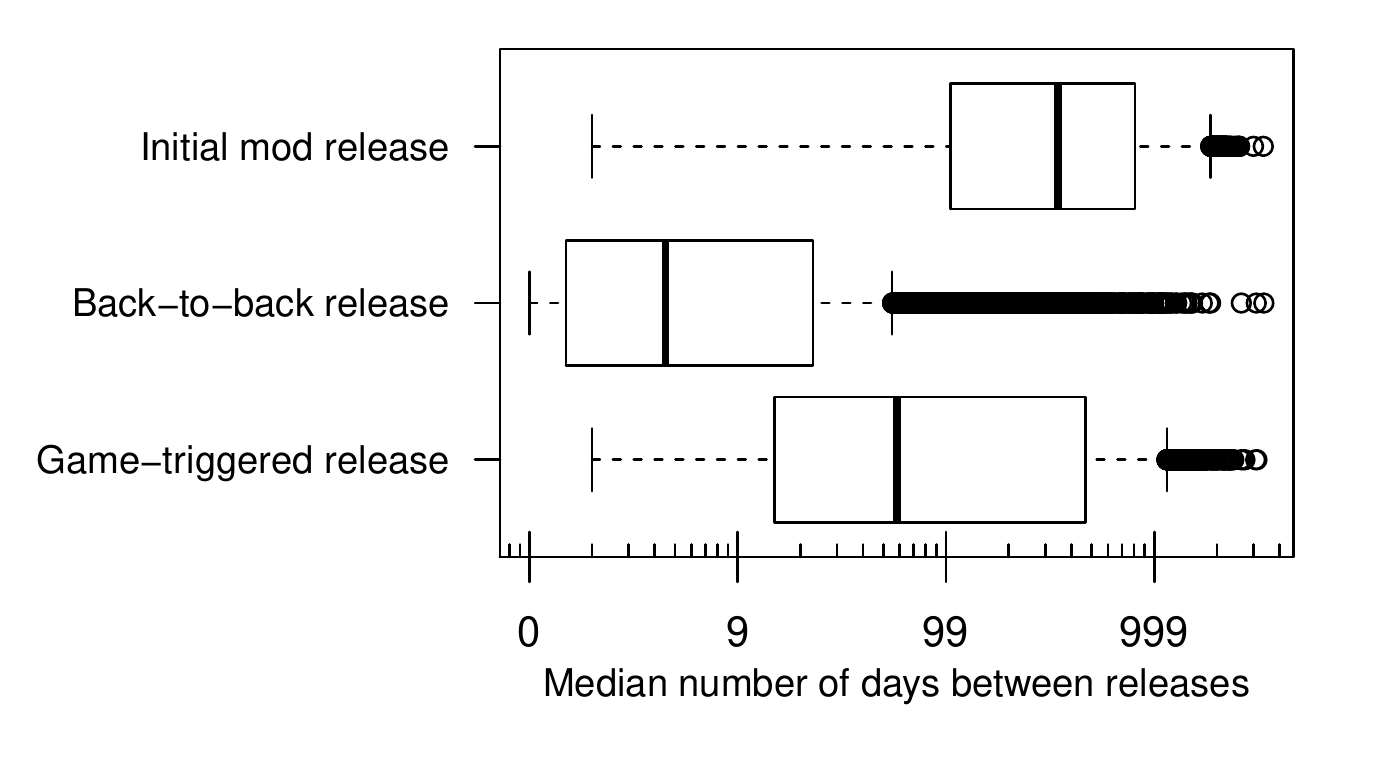}
  \caption{The distributions of the median number of days before a release for the three types of mod releases.}
  \label{fig:beanplot_logged_median_dist}
\end{figure*}

\textit{Findings:} \textbf{A mod has a median of two releases.} Figure~\ref{fig:multi_num_releases} shows the number of releases per mod across each studied dimension. The median number of releases is two across each studied dimension except for the non-media mods and mods of games without official modding support, which have a median of 3 releases. The Wilcoxon rank-sum test shows that we are unable to find a statistically significant difference under the significance threshold of $\alpha$ = 0.05 between the number of releases of mods for games with and without official modding support. For the media and non-media mods, the difference is significant but of negligible effect size. The finding for media and non-media mods is surprising, as one would expect that non-media mods would require more maintenance due to their closer relation with the original game's source code.

\textbf{There is a much longer time period between an official game release and a mod release, than between back-to-back mod releases.} 
Figure~\ref{fig:beanplot_logged_median_dist} shows the distributions of the median number of days between releases for the three types of mod release. We identified 9,478 initial mod releases, 32,207 back-to-back mod releases and 26,400 game-triggered mod releases. 
Back-to-back releases are by far the fastest releases, with a median of 3.5 days between releases. Since the median of the number of mod releases is two, a possible explanation for the fast back-to-back releases is that mod developers fine-tune (or fix) their mods based on feedback from the community. 

\textbf{It takes a median of 345.5 days before the initial version of a mod is released.} Figure~\ref{fig:beanplot_logged_median_dist} shows that the majority of mods are released after more than 345.5 days. These slow release times indicate that mod developers are either long-time players of a game, or have renewed interest in the game after a long time. Either way, such a lag in initial release indicates that mods are a way in which players of the game extend the longevity of a game. 

\textbf{Performance mods are released approximately ten times faster than other types of mods.} Table~\ref{tab:category_initial_release} shows the median number of days before a release for the three types of mod releases across the studied mod categories. Performance mods are released the fastest by far. However, we observed that performance mods are also the least endorsed, with a median endorsement ratio of 0.06. We found that the 11 performance mods in our data set are optimizations to improve the game's frames per second. However, several of these mods turned out to be counterproductive and actually decrease performance~\cite{performanceharmful}. In addition, the optimizations appear to target very specific combinations of hardware, which may leave users of the mod disappointed when the mod does not work for their hardware combinations.

\textbf{Save game mods take the longest to release after an official game release.} Table~\ref{tab:category_initial_release} shows that the initial release of a save game mod takes a median of 638.5 days. Save game mods consist often of saved games which allow the player to load a very advanced state of the game. For example, the \emph{Legit Skyrim Vanilla 99-Percent Completed} save game mod for the \emph{Skyrim} game has completed and collected 99\% of the quests and items in the game.
The game was played for a long time (more than 110 hours) to reach this advanced state. 
We observed that the median number of releases for save game mods was one. Save game mods with more than one release offered saved games for various story lines. For example, each release of the \emph{100 Percent Clean Save for Skyrim}\footnote{\url{https://www.nexusmods.com/skyrim/mods/56086}} mod offers a save game for a different story line in the game.

\begin{table}[!t]
	\caption{The median number of days between an official game update and a mod release for each mod category, along with the median number of days between adjacent mod releases, sorted by the median number of days before an initial mod release.}
	\centering
	\resizebox{\textwidth}{!}{
	\begin{tabular}{lrrr}
		\toprule
		& \multicolumn{3}{c}{\textbf{Median number of days before a release}} \\ 
		\multicolumn{1}{c}{\textbf{Mod Category}} & \multicolumn{1}{c}{\textbf{Initial mod release}} & \multicolumn{1}{c}{\textbf{Back-to-back mod release}} & \multicolumn{1}{c}{\textbf{Game-triggered mod release}}  \\
		\midrule
		Save Games  & 638.5 & 14.0 & 216.0 \\
		Bug Fixes  & 614.0 & 4.0 & 182.5\\
		Overhaul & 588.5 & 6.0 & 90.0\\
		Story  & 512.0 & 10.0 & 214.0\\
		Character & 476.5 & 3.5 & 127.0\\
		World Object & 442.0 & 6.0 &  81.3\\
		Utility  & 428.0 & 8.0 & 95.5\\
		Game Items & 339.0 & 2.0 & 49.0\\
		User Interface & 334.0 & 4.0 & 41.0\\ 
		Game Visuals  & 314.0 & 1.0 & 56.0\\
		Appearance & 309.0 & 2.0 & 67.5\\ 
		Gameplay & 274.5 & 5.0 & 50.0\\
		Audio & 194.5 & 6.0 & 37.0\\
		Cheats & 130.0 & 3.8 & 18.8\\
		Performance & 13.0 & 2.5 & 11.0\\
		\bottomrule
	\end{tabular}}
	\label{tab:category_initial_release}
\end{table}

\begin{figure}[!ht]
	\centering
	\subfloat[The \emph{Dragon Age} game franchise.]{
		\includegraphics[width=.5\textwidth]{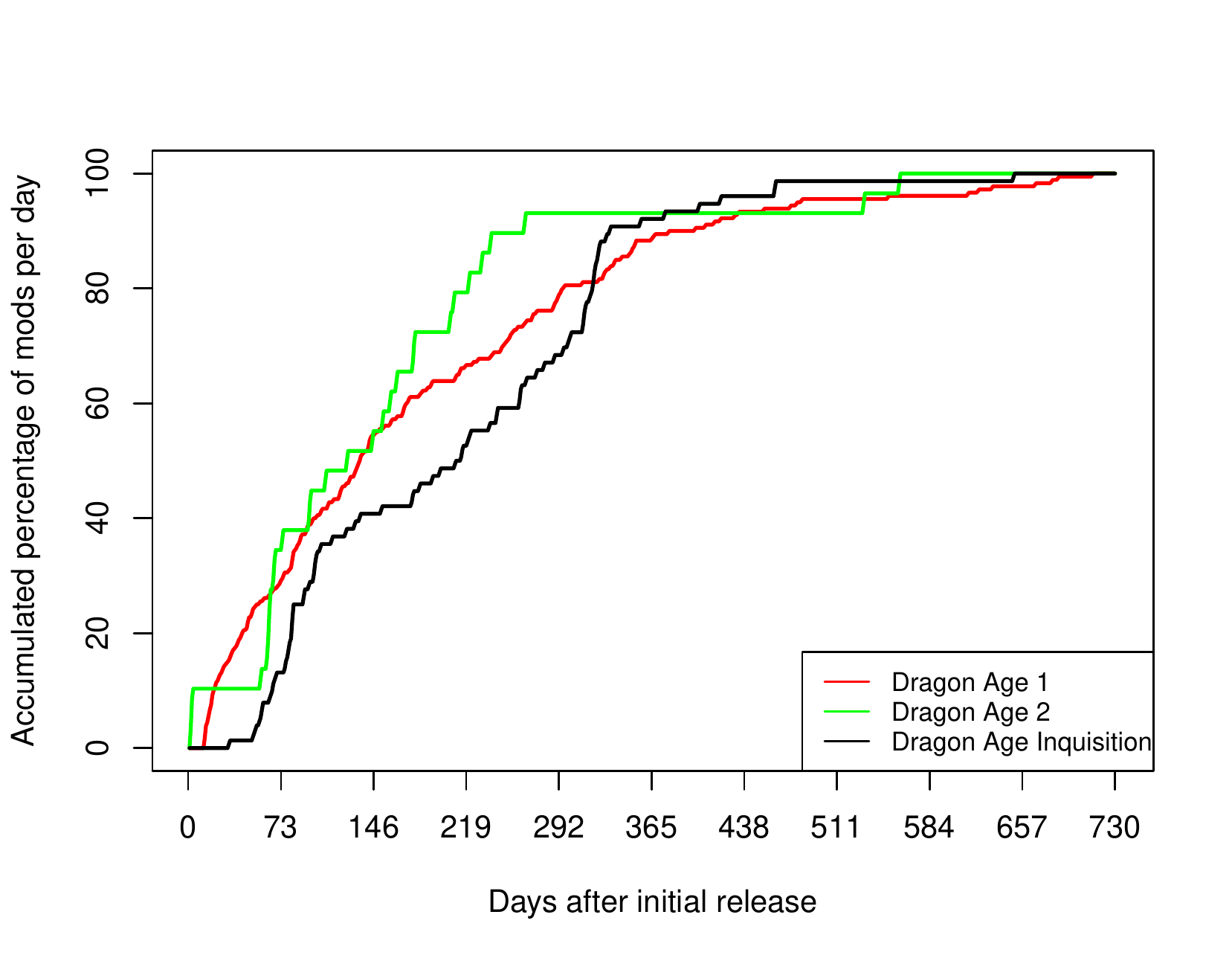}
		\label{fig:dragonage}
	}
	\subfloat[The \emph{Fallout} and \emph{Skyrim} game franchises]{
		\includegraphics[width=.5\textwidth]{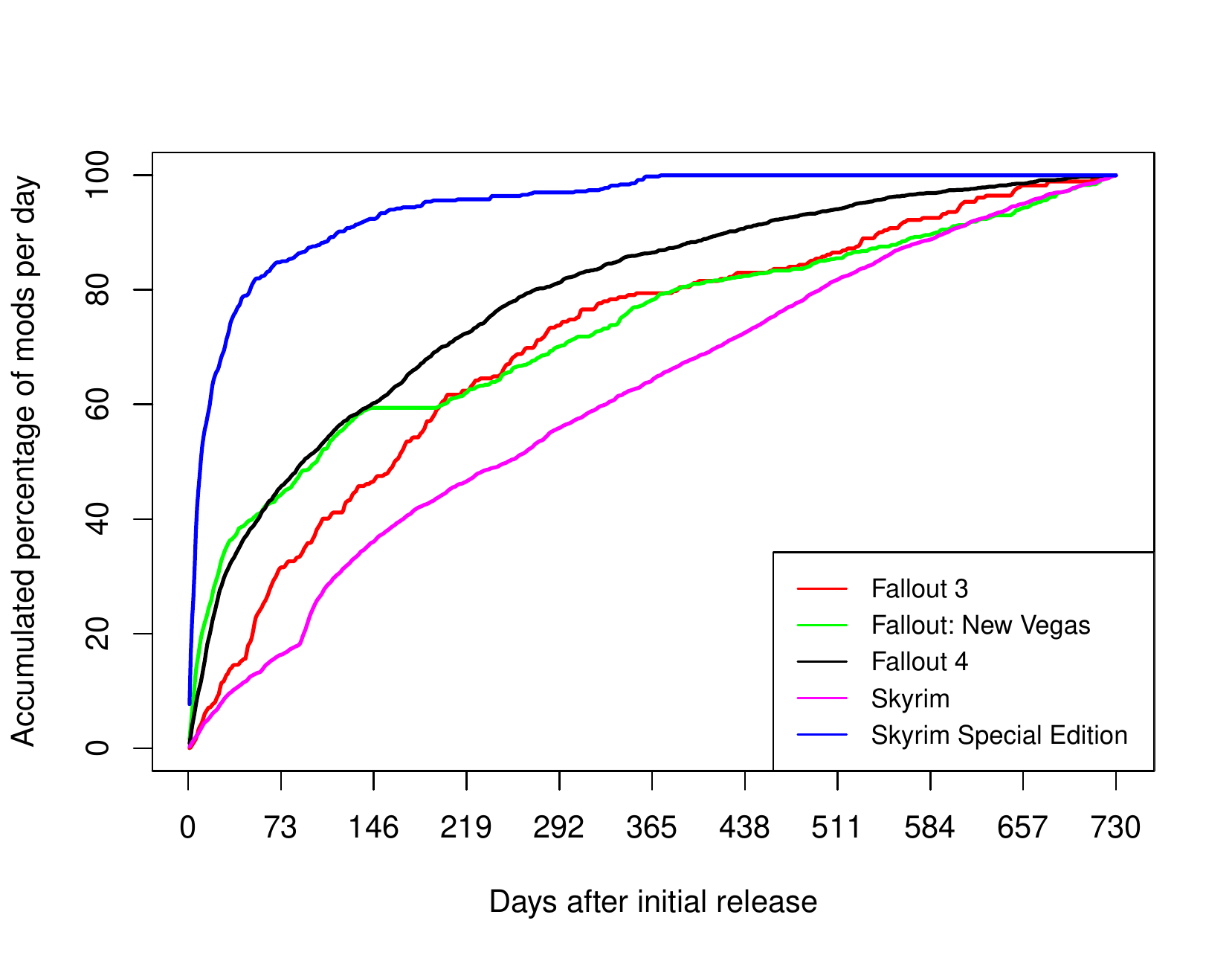}
		\label{fig:fallout}
	}\\
	\subfloat[The \emph{Witcher} game franchise]{
		\includegraphics[width=.5\textwidth]{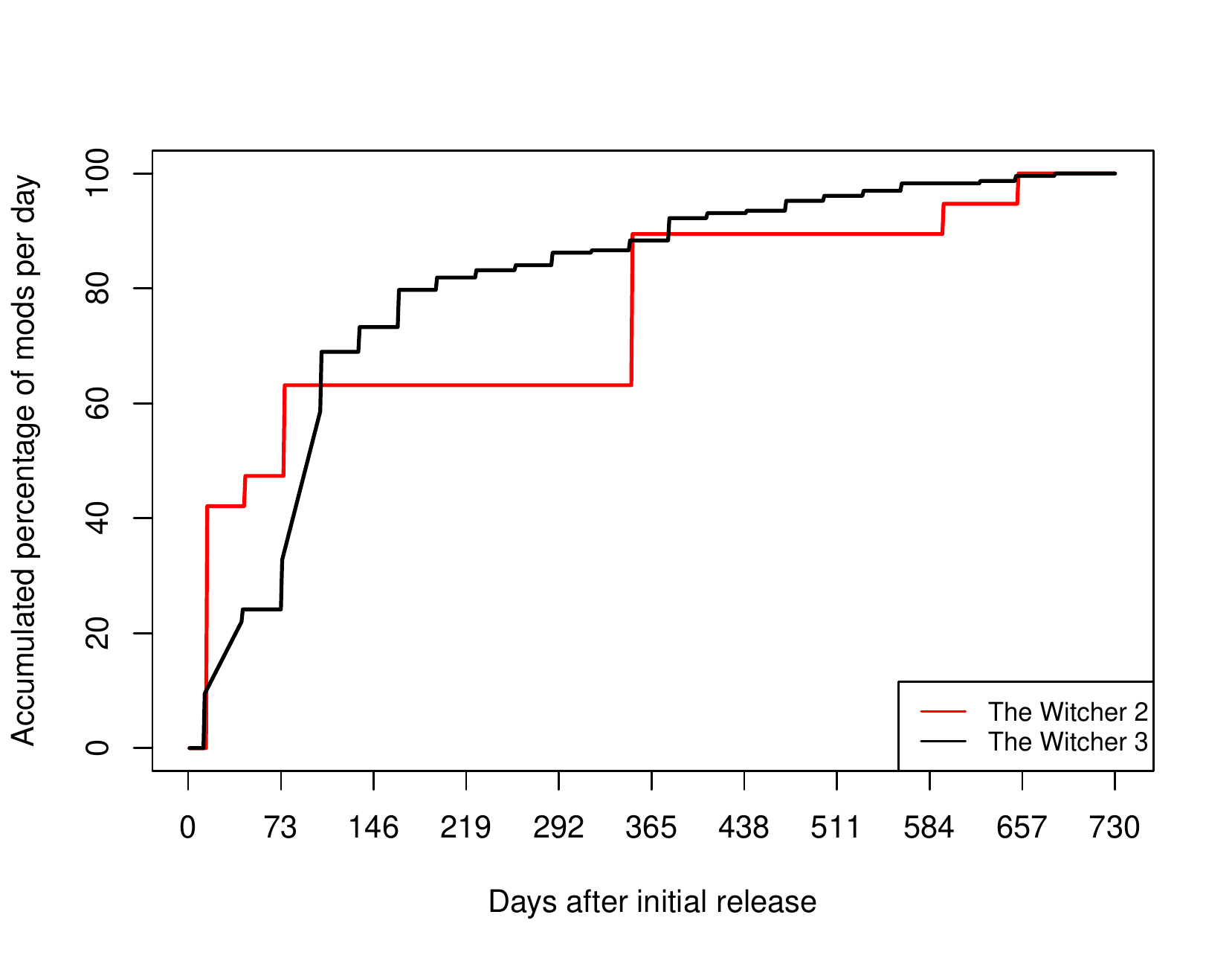}
		\label{fig:witcher}
	}
	\subfloat[The \emph{XCOM} game franchise]{
		\includegraphics[width=.5\textwidth]{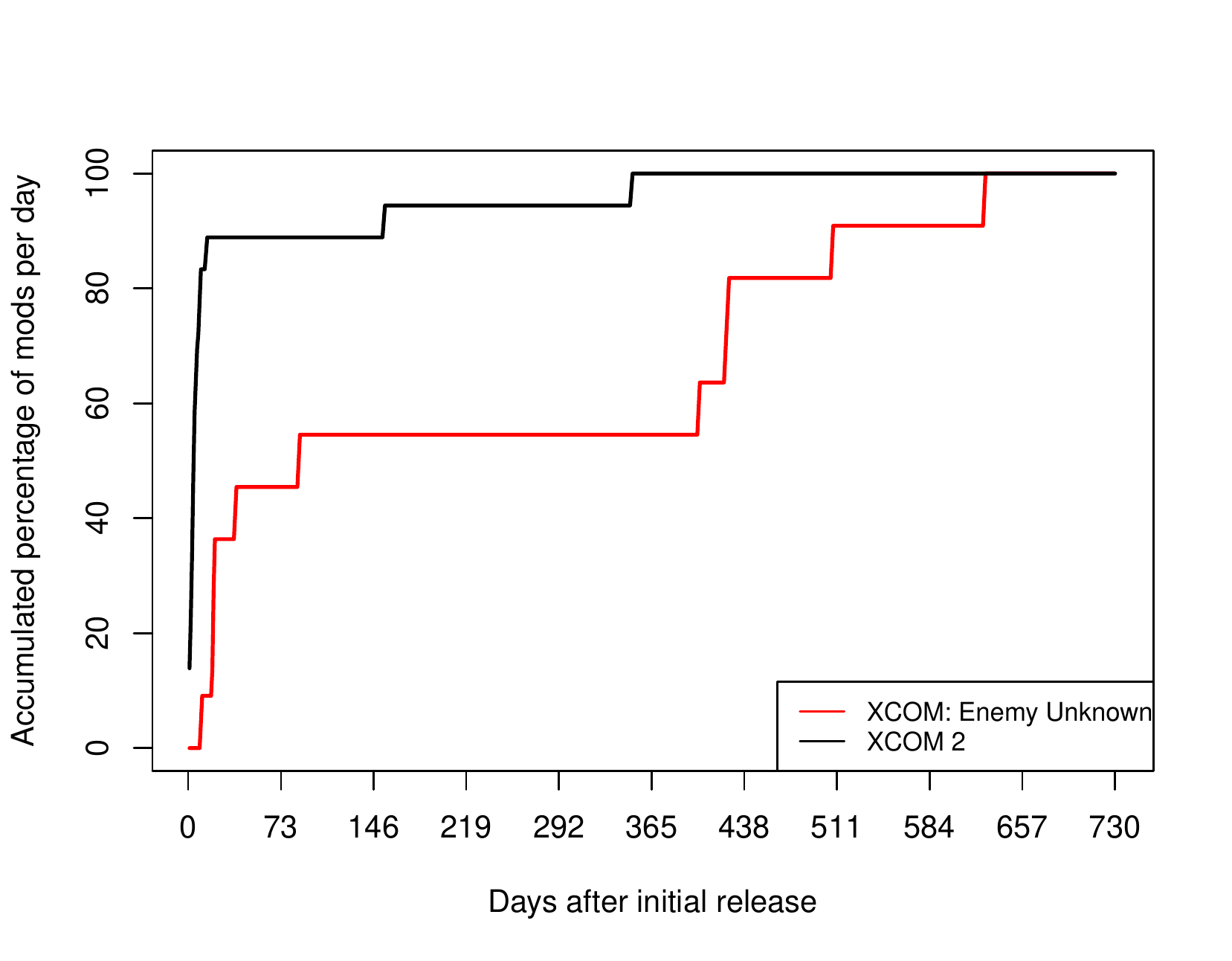}
		\label{fig:xcom}
	}\\
	\caption{The accumulated percentage of the number of released mods per day after the initial game release for the four studied game franchises. }
	\label{fig:lineplot_accumulated_perc}
\end{figure}

\textbf{The mods of games with official modding support are released faster than those of games without such support.} Figure~\ref{fig:lineplot_accumulated_perc} shows the accumulated percentage of mods that were released on a particular day after the initial game release for our four studied game franchises. Hence, a steep line indicates that mods are released relatively fast. Figure~\ref{fig:dragonage} shows that the mods for the \emph{Dragon Age 1} and \emph{Dragon Age 2} games are released faster than those of the \emph{Dragon Age: Inquisition} game, which was released after the \emph{Dragon Age 2} game. In particular, it took longer before the first mod of the \emph{Dragon Age: Inquisition} game was released: 32 days compared to 13 and 2 days for the \emph{Dragon Age 1} and \emph{Dragon Age 2} games. In addition, Figure~\ref{fig:xcom} shows that the mods for the \emph{XCOM 2} game were released much faster than for the \emph{XCOM: Enemy Unknown} game. Both the \emph{Dragon Age: Inquisition} and \emph{XCOM: Enemy Unknown} games do not have an officially-supported modding tool. Hence, Figures~\ref{fig:dragonage} and~\ref{fig:xcom} suggest that the mods of games with official modding support are released faster. A possible explanation is that it is likely easier to create such mods. 

\textbf{Supporting the same modding tool across games is associated with faster mod releases.}
Figure~\ref{fig:fallout} shows that the mods of the \emph{Fallout: New Vegas} and \emph{Fallout 4} games were both released faster than those of the \emph{Fallout 3} game. For the mods of \emph{Fallout: New Vegas} game, a possible explanation is that this game supports the same modding tool as the \emph{Fallout 3} game. While the \emph{Fallout 4} game supports a different modding tool, this tool was already supported by the \emph{Skyrim} game. Figure~\ref{fig:fallout} also shows that \emph{Fallout 4} mods were released much faster than \emph{Skyrim} mods. Figure~\ref{fig:fallout} suggests that when mod developers have the opportunity to get used to a modding tool, it allows them to release mods faster, or even convert mods to the later game. For example, the \emph{Asharas Hair Conversions} mod for the \emph{Fallout: New Vegas} game was ported from a mod for the \emph{Fallout 3} game. The impact of supporting the same modding tool across games is also demonstrated by the \emph{Fallout 76} game, which was released after we collected the data for our study. Although the \emph{Fallout 76} game would have official modding support at some point, modders did not wait for the official support and started to release mods for the game using the modding tool of the preceding \emph{Fallout 4} game~\cite{fallout76}.

A similar observation as for the \emph{Fallout} game franchise can be made in Figure~\ref{fig:dragonage} for the \emph{Dragon Age 1} and \emph{2} games, which support the same modding tool. In particular, the first 10\% of the \emph{Dragon Age 2} were released within four days which suggests that they could be converted with relatively low effort from existing \emph{Dragon Age 1} mods. Finally, an outlier in Figure~\ref{fig:lineplot_accumulated_perc} is the \emph{Skyrim Special Edition} game for which 42\% of the mods were released in the first week after its release, as the \emph{Skyrim Special Edition} game is mostly a graphical update of the \emph{Skyrim} game and hence shares most of the code base, making the mods easy to convert~\cite{skyrim_vs_skyrimse}.

\hypobox{\textit{A mod has a median of two releases, which are made closely after each other. Performance mods are by far the fastest released type of mod, but they often are not received well. The mods of games with official modding support are released faster than mods of games without such support. In particular, supporting the same modding tool across games within a franchise, or even different franchises, is associated with faster mod releases.}}

\begin{table}[!t]
\caption{The definitions of the bug statuses (sorted alphabetically).}
\centering
\begin{tabular}{ll}
\toprule
\multicolumn{1}{c}{\textbf{Bug Status}} & 
\multicolumn{1}{c}{\textbf{Definition}} 
\\
\midrule
Being looked at & The reported bug is currently under investigation.\\ 
Duplicate & The report is a duplicate of another bug report.\\
Fixed  & The reported bug is fixed.\\
Known issue & The reported bug was already known.\\ 
Needs more info & The bug report does not have sufficient information to fix the bug.\\
New issue & The bug report is new.\\
Not a bug & The reported bug is not reproducible or not a bug. \\
Won't fix & The reported bug will not be fixed.\\ 
\bottomrule
\end{tabular}
\label{tab:bug_status_definition}
\end{table}

\section{RQ2: How Well is the Post-Release Support of Mods?}
\label{sec:rq2}

\begin{figure*}[!t]
	\center
	\includegraphics[width=0.75\textwidth]{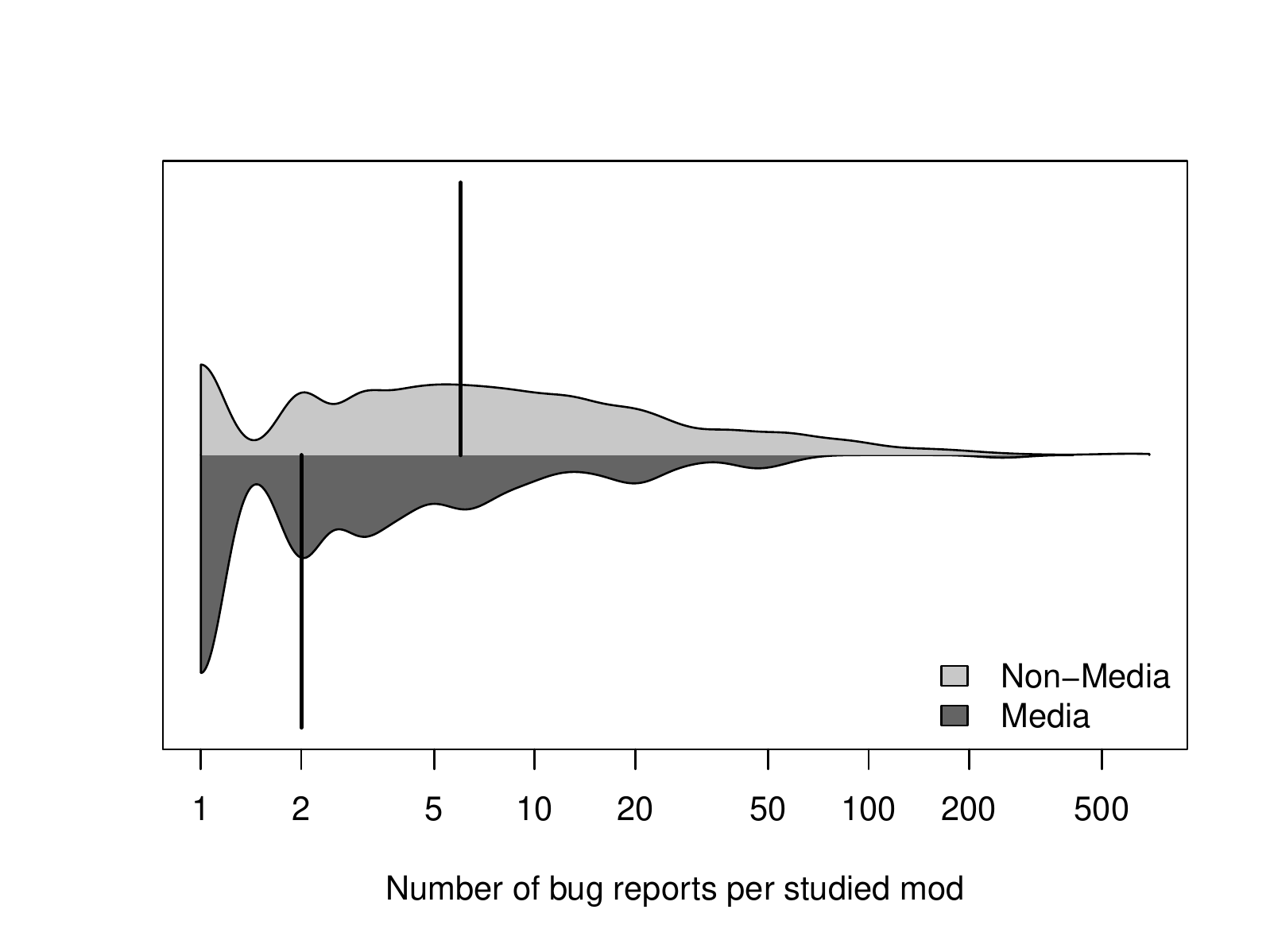}
	\caption{The distribution of the number of bug reports per studied non-media and media mod.}
	\label{fig:beanplot_num_of_bugs_media_vs_nonmedia}
\end{figure*}

\noindent\textit{Motivation:} In the previous section, we studied the release schedules of mods to learn about the willingness and speed of mod developers to release new versions of a mod. In this section, we study the willingness of mod developers to provide post-release support for their mods. In particular, we study how, and how fast mod developers deal with user-submitted bug reports. 

\textit{Approach:} To study how mod developers deal with bug reports, we studied the bug reports of the mods along several dimensions:
\begin{itemize}
	\item \textbf{The percentage of bug reports per bug status.} Table~\ref{tab:bug_status_definition} shows the bug report statuses that are supported by Nexus Mods. We calculated the percentage of bug reports per status for each studied mod. We filtered out 4,741 mods that had less than 10 bug reports as these mods with a low number of bug reports could bias our results in this dimension.	
	\item \textbf{The quality of a bug report.} We use the following features that were proposed by Bettenburg et al. in their study on what makes a good bug report~\cite{bettenburg2008makes}: the number of itemizations, the number of code samples, the number of stack traces, 	and the number of screenshots. In addition, we studied the length (in characters) of the bug reports and their comments per studied mod. To investigate the quality of the bug reports the first and third author independently manually analyzed a sample of 96 non-media and 90 media bug reports (which statistically represents the population with a 95\% confidence level and a 10\% confidence interval). 
	There were conflicting observations for two reports. One of the conflicts was due a confusion about what is a code sample, and one conflict was due to a confusion about what is a stack trace. We decided to use a broad definition of code sample and stack trace. Hence, we considered a snippet of a configuration file as a code sample, and logging output as a stack trace. After broadening our definition of code samples and stack traces, there were no further conflicts between the authors' classifications.
	\item \textbf{The people working on the bug report.} We studied how many and which people were involved in the bug report (i.e., the poster and the people commenting on the report). 
\end{itemize}

\begin{figure*}[!t]
	\center
	\includegraphics[width=0.75\textwidth]{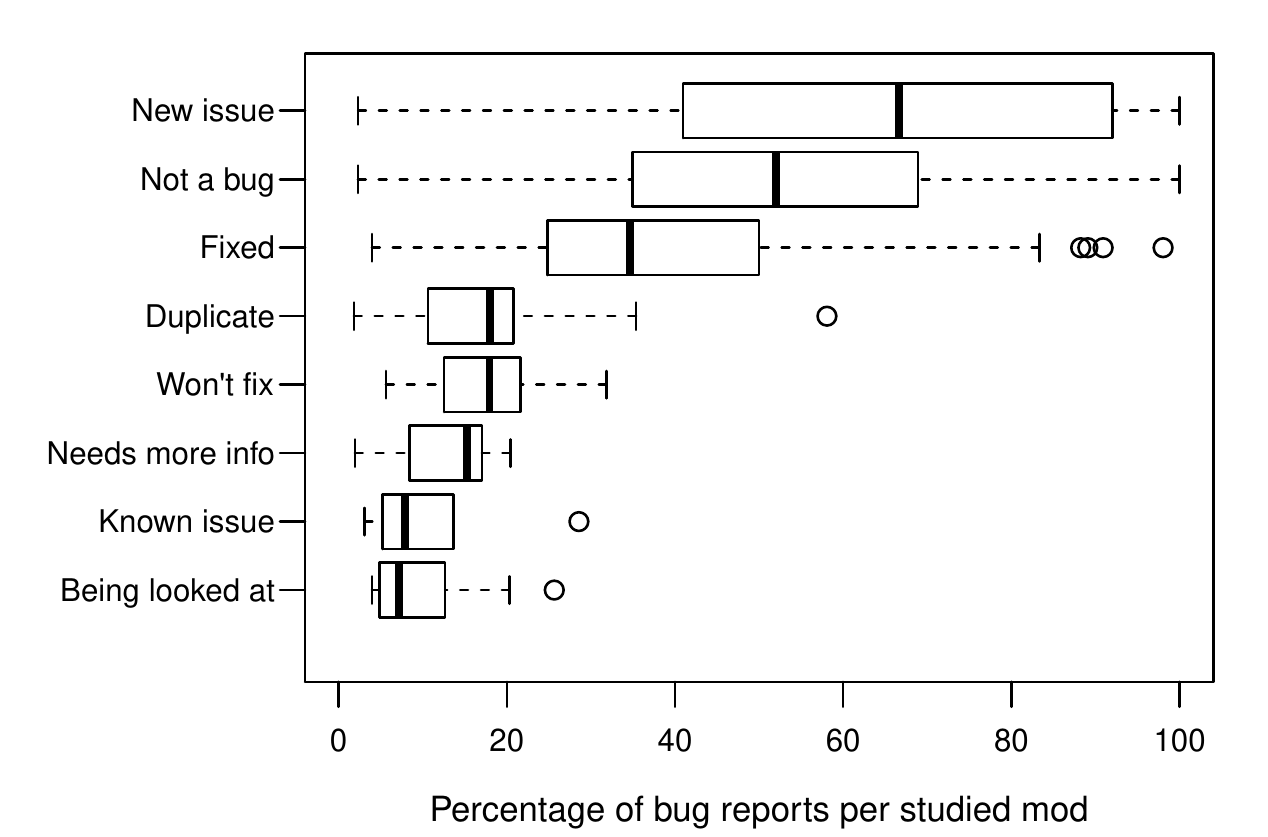}
	\caption{The distribution of the percentage of bug reports per studied mod for each bug status. Sorted by the median percentage of bug reports per studied mod.}
	\label{fig:boxplot_num_bug_status_per_mod}
\end{figure*}

\begin{figure*}[!t]
\center
    \includegraphics[width=0.75\textwidth]{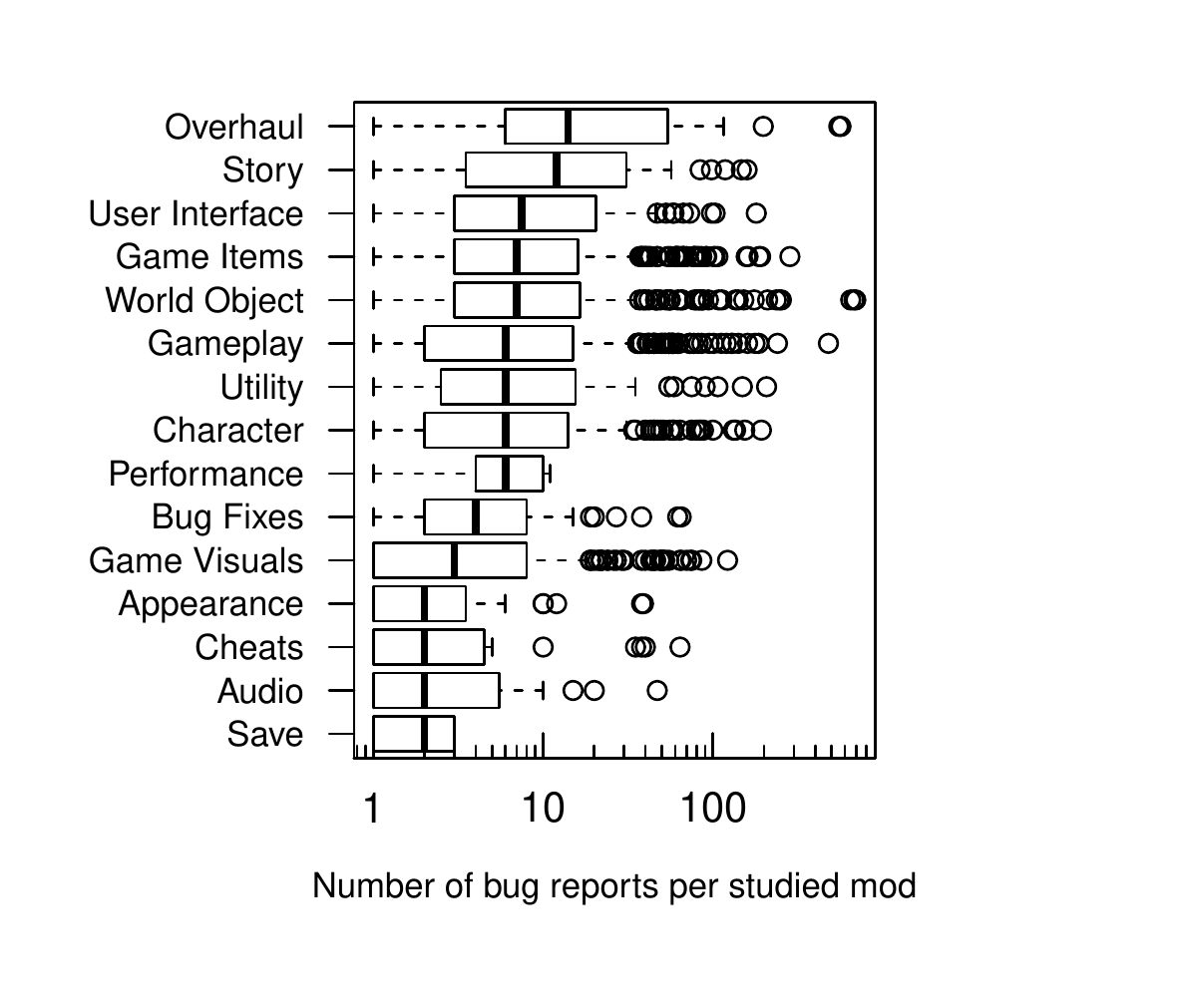}
  \caption{The distribution of the number of bug reports per studied mod for each high level mod category.}
  \label{fig:boxplot_number_of_bug_reports_per_mod}
\end{figure*}

\textit{Findings: }
\textbf{Non-media mods have a higher number of bug reports per studied mod than media mods.} Figure~\ref{fig:beanplot_num_of_bugs_media_vs_nonmedia} shows the distributions of the number of bug reports per studied non-media and media mod. The Wilcoxon rank-sum test shows a statistically significant difference between the number of bug reports per mod for non-media and media mods, with a medium Cliff's delta effect size. Non-media mods have a median number of six bug reports per studied mod and media mods have a median number of two bug reports per studied mod. An intuitive explanation is that non-media mods are more complex than media mods, and are therefore more error-prone. We are unable to find a statistically significant difference under the significance threshold of $\alpha$ = 0.05 between the number of bug reports for mods with and without official modding support, and mods for Bethesda and non-Bethesda games.

\textbf{Complete overhaul mods have the highest median number of bug reports per mod.} Figure~\ref{fig:boxplot_number_of_bug_reports_per_mod} shows the number of bug reports per studied mod for all mods across each high level mod category. The median number of bug reports per mod for overhauls is 14. However, the number of bug reports per mod across overhauls varies greatly. 
We manually sampled 39 overhaul mods and found that 26 of the 39 overhauls mods (67\%) with bug reports added new content to a game. The median number of bug reports for these 26 mods was 16.

\textbf{A median of 67\% of the bug reports of a mod have the `New issue' status.}
Figure~\ref{fig:boxplot_num_bug_status_per_mod} shows the percentage of bug reports per studied mod for each bug status. For most studied mods, the majority of the bug reports are labeled with the `New issue' status. 
Prior work~\cite{copingwithopenbug} on bug tracking systems of the Eclipse and Firefox open source projects reported that 18-33\% of the bug reports have a status that is equivalent to the `New issue' status on Nexus Mods. The percentage that we found is considerably higher. There are several possible explanations for this difference. First, there is a large difference in scale between the studied open source projects in prior work and our studied mods. In particular, many of these mods were created by a single developer who may lack the time or simply the interest to fix all reported bugs. Second, we observed during our manual study that the vast majority of the bug reports were of low quality, making them difficult to interpret and address. A possible explanation is that gamers find it difficult to report stack traces or code samples, since it is commonly not trivial to retrieve them (e.g., in games where it is difficult to cut and paste a stack trace). Third, we learned during our manual analysis that many users have several mods for a game installed. When a new version of the game is released, often one or more of these mods break. Unfortunately, it is hard for users to identify which of the mods are broken and therefore, they report all their installed mods as broken. For example, the developer of the popular \emph{90000 weight limit} mod for the \emph{The Witcher 3} game explains that while most game releases break the mod, the mod is always updated.\footnote{\url{https://www.nexusmods.com/witcher3/mods/836?tab=posts}} 
The developer explains that many of the reported bugs are not related to the mod and he gives several suggestions on how to rule out problems with other mods before submitting a bug report.

\textbf{The vast majority of bug reports for mods are of low quality.} Bettenburg et al.~\cite{bettenburg2008makes} showed that bug reports that contain stack traces get fixed sooner while bug reports that contain code samples have a higher likelihood of getting fixed. 
During our manual analysis we found that 3\% of the bug reports for non-media mods contained a stack trace, 2\% contained a code sample and 6\% contained a screenshot or video. 
In addition, 3\% of the bug reports for media mods contained a stack trace, 4\% contained a code sample and 4\% contained a screenshot.
Hence, according to Bettenburg et al.'s study it is not surprising that many of the bug reports on Nexus Mods are not yet fixed.
Interestingly, bug reports for media mods contained a code sample more often than bug reports for non-media mods. While the higher percentage of code samples for media mods is surprising, a manual inspection of the bug reports shows that most of these code samples are not actual source code, but a snippet of a configuration file. Such confirmation files are essential to activate mods.

\textbf{If a bug report gets a response, the median number of days to get that response is 0. Likewise, if a bug report gets fixed, this takes a median of 1 day.} While many bug reports do not get a response at all, the ones that got a response often received it on the same day as the bug was reported. In addition, while many bug reports remain unsolved, the ones that are solved are often solved within one day. Hence, it seems that mod developers are willing to solve and respond to bug reports quickly. A possible explanation is that mod developers would quickly address trivial issues, and avoid working on the more difficult ones. However, as explained above, the low quality of bug reports often prevents mod developers from addressing issues at all.

\begin{figure*}[!t]
\center
    \includegraphics[width=0.75\textwidth]{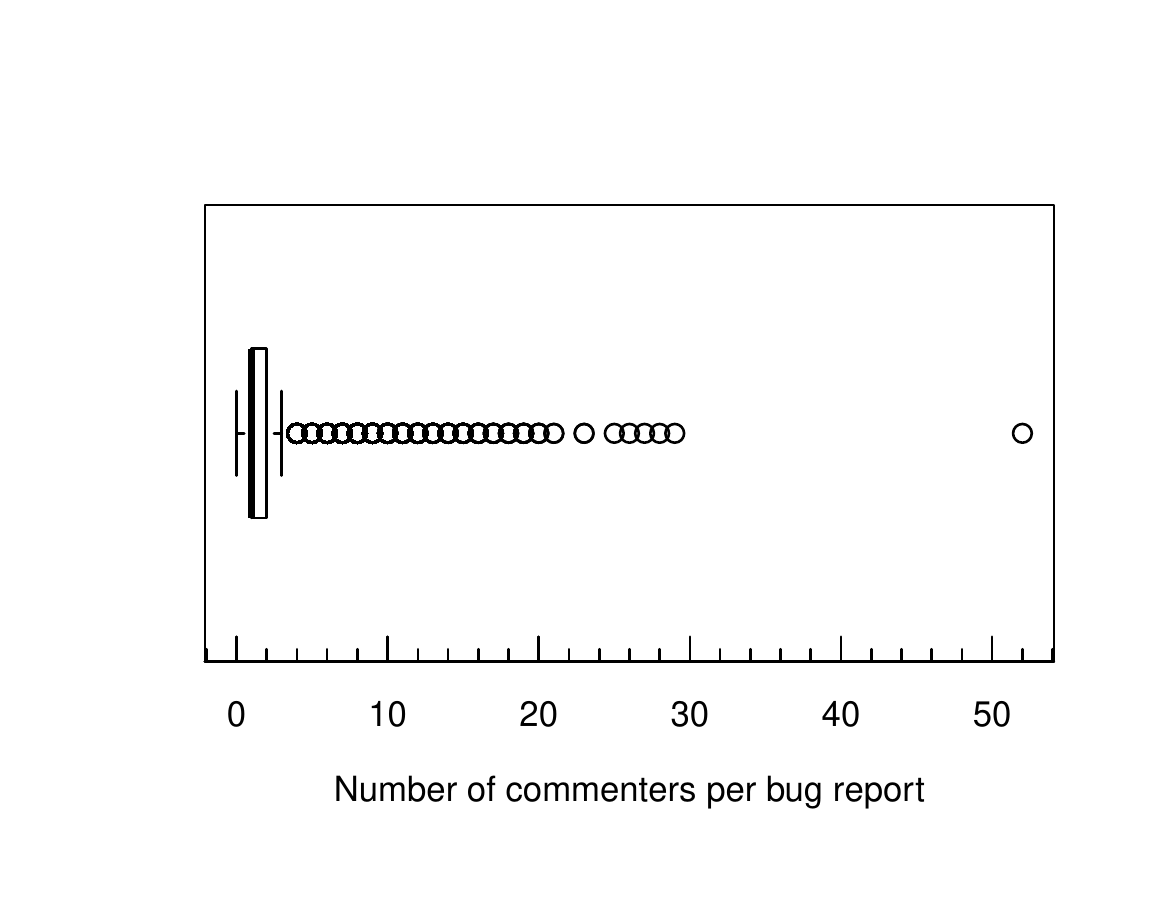}
  \caption{The distribution of the number of unique commenters per bug report.}
\label{fig:boxplot_num_distinct_commenters_per_bug}
\end{figure*}

\textbf{In 51\% of the bug reports, the first comment is by someone other than the mod developer.} Figure~\ref{fig:boxplot_num_distinct_commenters_per_bug} shows that the median number of commenters per bug report is one, which would suggest that the discussion is done between the bug reporter and the mod developer. However, we found that in 51\% of the bug reports with one comment, the discussion was actually between the bug reporter and another community member. This relatively high percentage indicates the engagement of the community of a mod. We did observe a difference between the percentage of fixed bug reports: of the 1,642 fixed bug reports with one comment, 66\% of the reports that received a comment from the mod developer were fixed compared to 34\% of the reports that received a comment from someone other than the developer. In the latter case, we observed several indications that the engaged community member was a contributor of the mod in question, or that the bug report was fixed without a message from the mod developer.

\begin{figure*}[!t]
\center
    \includegraphics[width=0.75\textwidth]{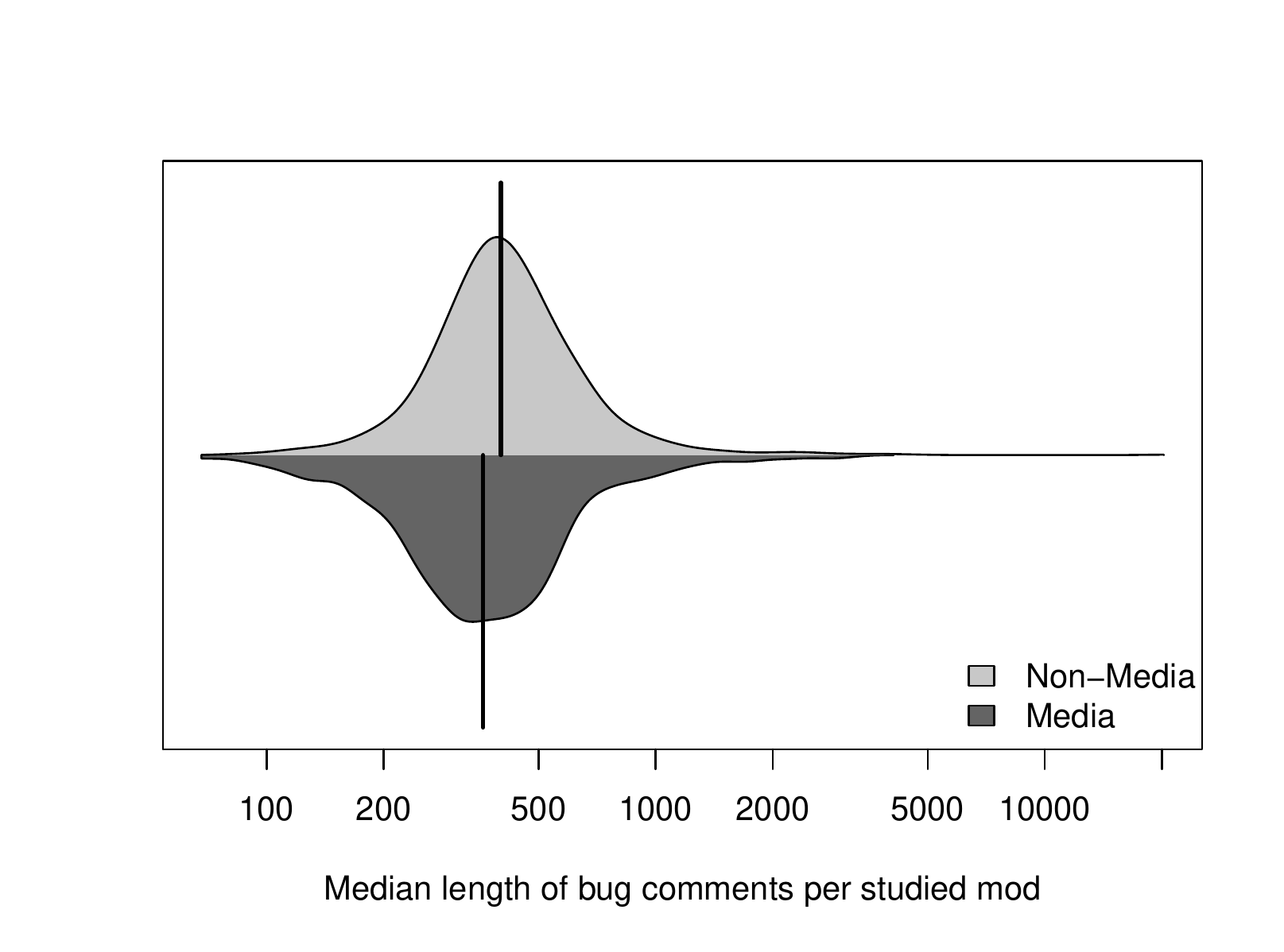}
  \caption{The distribution of the median length of the bug report comments per studied mod for non-media and media mods.}
  \label{fig:beanplot_compare_length_media_nonmedia}
\end{figure*}

\textbf{The bug report comments of non-media mods have a greater median length in characters than those of media mods.} Figure~\ref{fig:beanplot_compare_length_media_nonmedia} shows the median length in characters of bug report comments of non-media (400 characters) and media mods (360 characters). The Wilcoxon rank-sum test shows a significant difference between the median lengths, with a small Cliff's delta effect size.  
This difference suggests that bug reports for non-media mods are slightly more complex than those of media mods.

\hypobox{In general, bug reports for mods are of low quality. As a result, many of them remain unsolved. In addition, in many bug reports, community members other than the mod developers are engaging with the bug reporter.}

\section{Implications of our Findings}
\label{sec:implications}
In this section, we outline the most important implications of our findings for game developers and developers of mod distribution platforms. 

\subsection{Implications for Game Developers}
\textbf{Game developers who desire a modding community for their game should consider supporting an official modding tool.}
One of the goals of our paper is to investigate whether providing official modding support for a game contributes to a healthy modding community.
In our preliminary study we observed that 9 of the 10 most mod-popular games on the Nexus Mods platform have an official modding tool. In addition, we showed in RQ1 that the mods for games with official modding support are released faster than the mods for games without such support. In particular, this release speed improves for games that use the same modding tool as other games, which often allows mod developers to easily convert mods for newer games. An example of a game developer who does well at supporting the modding community is Bethesda. In fact, when Bethesda released the \emph{Skyrim Special Edition} game, it was compared by journalists not only to the original \emph{Skyrim} game, but also to the modded version of that game~\cite{skyrim_vs_skyrimse}. As Bethesda uses the modding community of their games as a breeding ground for hiring new developers~\cite{whowroteeldderscrolls}, they show that it can be beneficial for game developers to maintain a healthy modding community.

\textbf{Game developers should not only provide support for creating a mod, but also for maintaining it.} 
Most of the studied mods are not well maintained after their initial release. In RQ1, we observed that the median time between back-to-back mod releases is only 3.5 days. 
In addition, in RQ2 we observed that mod developers respond to and solve some bug reports very fast. Hence, there are several indications that mod developers and the modding community are willing to provide continuous maintenance for mods.
Hence, game developers who wish to support the modding community of their games should also consider providing better support for maintaining the mod. Therefore, game developers should continuously improve modding tools after a game release in order to help drive the popularity of the original game. For example, DOTA, Counter-Strike, and DOTA Auto Chess are mods within games who have helped drive the popularity of the original games.

\subsection{Implications for Developers of Mod Distribution Platforms}
\textbf{Mod distribution platforms should consider providing tools for detecting conflicts between mods.}
During our study, we often came across cases in which the mod developer provides an informal disclaimer about the compatibility of a mod with other mods of the game. In most of these cases, the mod developer does not guarantee compatibility. The main reason is that mods for the same game often modify the same files. Hence, there may be conflicts between these modifications, especially because many users install several mods for a game. Because of the large number of mods for some games, it is not feasible to test a mod in all possible installation configurations. We envision mod distribution platforms acting in a supporting role when it comes to detecting conflicts between mods. For example, these platforms could inspect the files of a mod that is being submitted and report potential conflicts with other mods. The mod developer can then focus on testing these combinations of mods.

\textbf{The Nexus Mod distribution platform should improve its bug reporting system, e.g., by guiding users better when submitting a bug report.}
We observed in our study that the vast majority of the bug reports for mods are of low quality. 
An important first step to improving the quality of the bug reports would be to better guide users when they are reporting a bug. For example, we observed that a very small portion of the bug reports contains a stack trace, code sample or screenshot. 
Explicitly asking the bug reporter to add such clarifications for a bug report may increase its quality. In addition, future studies should investigate what a good bug report for mods looks like. For example, we observed that mod developers ask users to post the loading order of their mods when submitting a bug report to investigate possible conflicts between mods.\footnote{\url{https://www.nexusmods.com/Core/Libs/Common/Widgets/ModBugReplyList?issue_id=18914}} 
To identify the percentage of bug reports that contain a loading order, we searched all bug reports for several variations of the ``load order'' keywords and found that only 4\% of the bug reports mentions these keywords. After manually studying a representative sample of 89 of those bug reports (95\% confidence level and 10\% confidence interval), we found that only 25 out of those 89 (28\%) bug reports actually contain a loading order. Hence, the total number of bug reports that contain an actual load order is very small (approximately 1\%). 
We envision that mod distribution platforms explicitly ask bug reporters to include their mod loading order.

In addition, we came across many complaints about bugs of mods in other sections than the bug reports. For example, in the `posts' section of the \emph{90000 weight limit} mod, there are many mentions of bugs, while the mod only has two bug reports. Hence, the current bug reporting tool is either unknown by or unclear to mod users.

\subsection{Implications for Mod Developers}

\textbf{Mod developers should make it easier for gamers to collect in-game stack traces or code samples.} In our study, we observed that bug reports tend to be of low quality. A possible explanation is that gamers find it difficult to report stack traces or code samples, since it is commonly not trivial to retrieve them, e.g., in games where it is difficult to simply cut and paste a stack trace. Hence, we suggest that mod developers provide gamers with an automated and simple mechanism for collecting in-game stack traces or error symptoms.

\section{Research Challenges}
\label{sec:challenges}

In this section, we highlight the most important research challenges that we encountered during our study. These challenges all require dedicated studies to be addressed.

\subsection{Establishing a Ground Truth for the Impact of a Modding Community}
A major challenge in research on game mods is that it is hard to establish a ground truth for the impact of a modding community on the original game. For example, it is difficult to estimate how the popularity of a mod-popular game would have evolved without its modding community. In our work, we made a first step towards estimating the impact of official modding tools on game mods by studying games within the same game franchise. While definitely not perfect, studying games within the same game franchise mitigates some of the confounding factors that exist when studying completely different games. Further studies are necessary to determine how a ground truth for the impact of a modding community can be determined. For example, future studies should conduct a large scale and longitudinal investigation of players of games with and without mods.

\subsection{Studying the Impact of Game Mods on the Life Expectancy of the Original Game}
Another potential benefit of game mods for game developers is that game mods could increase the life expectancy of the original games. In our data, we have observed several ``older'' games (such as Skyrim) which are still very popular, possibly because of the modding community. Future studies should investigate how the impact of mods on the life expectancy of a game can be measured reliably. First, one should decide how to measure life expectancy. For example, do we consider the number of total playing hours on Steam as a proxy for a game's life expectancy? The challenge of this metric is that it would be hard to associate with the existence of a modding community, as there is no way to verify whether a Steam gamer has modded the game, and whether the mods actually impact a game's life expectancy. In addition, there could be many confounding factors that impact the life expectancy of a game. For example, the life expectancy could be impacted by the type of the game, news coverage about the game, the addition of official new levels/items, an inherently more loyal playerbase, etc. These confounding factors cannot be ignored yet they are currently impossible to quantify in a reliable manner. A possible direction for future work is to collaborate with developers of games with an active modding community and monitor how many of the currently active users run their games with mods. In addition, such collaboration could also yield further insights on which mods are being used.

\subsection{Studying the Impact of Game Mods on the Ratings of the Original Game}
A research challenge that we uncovered in our study is the complexity of analyzing the relations between a modding community and ratings of its original game. Understanding how game mods (or the existence of a modding community) affect the rating of the original game would help understand the benefits of game mods for game developers. Unfortunately, there are many confounding factors that impact the ratings of a game. For example, the ratings can be impacted by good/bad updates~\cite{hassan_badupdates} or a change to the price of the game. Hence, there currently is no reliable mechanism to correlate the presence of a modding community with the original game's rating. In addition, the benefits of a modding community for game developers may be reflected in other metrics than the rating. For example, a game could have more daily players because of its mods. However, the impact of a modding community on such metrics is again difficult to reliably quantify, as there are many confounding factors.  Future studies should investigate reliable methods of quantifying the impact of the presence of a modding community on the ratings of a game to provide deeper insights on the benefits of game mods for game developers.

\subsection{Studying the Impact of Game Mods on the Time Required to Develop Features}
Another potential benefit of game mods is that they can be integrated into the original game, thereby potentially saving the implementation time. 
Unfortunately, this benefit is impossible to study at a large scale, since these games are not open source, making it impossible to access their source code or issue tracking systems. However, even with such access, the time to develop a new feature would be hard to quantify. The vast majority of new features in a game is not based on mods, and likewise the vast majority of game mods will never end up in the original game. Future studies that have access to development team or source code history data should use qualitative methods such as interviews to observe if and how mods impact the development time of a feature.

\section{Threats to Validity}
\label{sec:threats}
This section outlines the threats to the validity of our findings.

\subsection{Internal Validity}
A threat to the internal validity of our study is the way in which our studied dimensions were defined. For example, we do not know which portion of the mods for a game with official modding support were actually built using the officially-supported tool. However, we use this dimension to indicate the mindset of a game developer rather than draw conclusions about the quality of the modding tool or the mods that are created with it. Future studies should further investigate whether the quality of mods that are created using an officially-supported tool is higher.
In addition, we chose a threshold to distinguish media and non-media mods. This threshold was selected based on the distribution of the portion of media files in a mod, and the threshold takes into account the number of files in a mod. 
To mitigate this threat, we repeated our analysis after lowering the thresholds to 90\% and 85\%, and we found that our findings still hold.

Another threat to the internal validity of our study is that game developers are not required to publish official game release notes~\cite{urgentupdates}. Hence, we may have missed game releases which in turn affects our calculation of the release time of a mod. In addition, as the data on the Nexus Mods platform is barely moderated, it is quite noisy and we have no guarantee that the publication date of a mod is its release date. Finally, not all releases of a mod may end up on Nexus Mods. For example, game developers may release a closed beta-version of their game. Hence the numbers that we report in our study are all low bound estimates.

Another threat is that we use the release schedule of mods as a proxy for the willingness of mod developers to release new versions of a mod. It is possible that a mod requires only one release. For example, simpler mods, such as texture mods, could be perfect after their initial release (even though we observed in Section~\ref{sec:rq1} that media mods have a median of two releases as well). To verify the impact of this threat, 
we attempted to identify whether single-release mods were indeed perfect. First, we observed that 74\% of the bug reports that were posted after the single release have the `New issue' status. This 74\% is higher than the overall percentage of mods with the `New issue' status (67\%). Hence, this higher percentage is an indication that often mod developers of single-release mods abandoned the mod while it still required maintenance, rather than that the mod was perfect. In addition, we manually studied a sample of 50 bug reports that were posted after the release of a single-release mod, and we observed that 13 of those bug reports were confirmed by others. Hence, the reported bugs affects multiple modders, making it more likely that the mod requires a bug fix (and is not perfect).

Finally, it is important to keep in mind that the actions of the original game developers may bias our findings. For example, the original game developers could decide to integrate a mod in the original game, or they could ensure the proper functioning of the mod across releases of the original game. As a result, the mod may appear unsupported on Nexus Mods. Future studies should further investigate how game developers react to mods.

\subsection{External Validity}
A threat to the external validity of our study is that we only studied mods from the 20 most mod-popular games on the Nexus Mods distribution platform. As a result, we did not study mods for games that may be mod-popular, but do not have a strong presence on the Nexus Mods platform. An example of such a game is the \emph{Minecraft} game, which is popular amongst modders but of which the mods are distributed through other platforms, such as the \emph{Planet Minecraft} platform.\footnote{\url{https://www.planetminecraft.com/resources/mods/}}
Future studies should investigate how our findings apply to games and mods that are available through other mod distribution platforms.

Another threat concerning the generalizability of our study is that we only studied mods with at least 1,000 endorsements. We conducted a sensitivity analysis to see if our findings are biased by the 1,000 endorsement threshold. We repeated our analysis using 750 and 1,250 as endorsement thresholds and observed that our findings still hold. Future studies should investigate whether our findings apply to less popular mods as well.

\section{Conclusion}
\label{sec:conclusion}

Game mods are community-driven contributions to a game that are made by players of the game who wish to improve existing or add new functionality to the game. For example, a mod can improve the textures of a game, add new weapons to the game, or even provide a complete overhaul of the game.
In this paper, we studied 9,521 popular mods from the 20 most-modded games on the Nexus Mods distribution platform. In particular, we studied the characteristics, release schedules and bug reports of the mods, to get a better understanding of the willingness of the modding community of a game to contribute and maintain mods. We studied mods along three dimensions: (1)~mods for games with and without official modding support, (2)~media and non-media mods, and (3)~mods of Bethesda and non-Bethesda games. Below, we list our most important findings:

\begin{enumerate}
\item Mods for games with official modding support have a higher median endorsement ratio than mods for games with no official modding support. In addition, mods for games with official modding support are released faster.
\item Supporting the same modding tool across games within a game franchise, or even different franchises, is associated with faster mod releases.
\item A mod has a median of two releases, which are often released within four days of each other. In addition, mod developers respond to and solve bug reports quickly.
\item Unfortunately, bug reports for mods are of low quality, which make them hard to address. In addition, many bugs are reported outside the bug reporting tool.
\end{enumerate}

In this paper, we found several indications that the modding communities of games are willing to contribute to mods. Many of these mods are being used by hundreds of thousands of gamers. Hence, while we are definitely not suggesting that game developers should rely on mod developers for the core functionality of their games, mods are a good and cost-effective way of keeping the players of a game engaged. As a result, it can be beneficial for game developers to provide official support for modding their games.
We suggest that game developers and developers of third-party mod distribution platforms step in to help mod developers improve the post-release support of their mods. Game developers could offer incentives to mod developers who provide excellent post-release support for their mods. For example, popular mods that are well-maintained could be included in an official release of the game. In addition, developers of mod distribution platforms could improve the infrastructure for reporting and addressing bugs, thereby striving to improve the quality of the bug reports.

\bibliographystyle{spbasic}      
\bibliography{biblio}

\end{document}